\documentclass[aps,groupedaddress]{revtex4}
\usepackage{epsfig}
\usepackage{graphicx}
\usepackage[T1]{fontenc}
\usepackage{ae}
\usepackage{color}
\usepackage[latin1]{inputenc}
\usepackage{amssymb,amsbsy,amsmath}
\usepackage{bbm}
\begin{document}

%\begin{center}\today\end{center}

%%%%%%%%%%%%%%%%%%%%%%%%%%%%%%%%%%%%%%%%%%%%%%%%%%%%%%%%%%%%%%%%%%%%%%%%%%%%%

\title[Theory of ground states VI]
{Theory of ground states for classical Heisenberg spin
systems VI}

\author{Heinz-J\"urgen Schmidt$^1$ and Wojciech Florek$^2$
}
\address{$^1$  Universit\"at Osnabr\"uck,
Fachbereich Physik,
 D - 49069 Osnabr\"uck, Germany\\
$^2$  Adam Mickiewicz University, Faculty of Physics,
ul.~Uniwersytetu Pozna\'{n}skiego 2, 61-614 Pozna\'{n}, Poland}

%\tableofcontents

\begin{abstract}
We formulate part VI of a rigorous theory of ground states for classical, finite, Heisenberg spin systems.
After recapitulating the central results of the parts I - V previously published we consider a magnetic field
and analytically calculate the susceptibility at the saturation point.
To this end we have to distinguish between parabolic and non-parabolic systems, and for the latter ones
between two- and three-dimensional ground states.
These results are checked for a couple of examples.
\end{abstract}

\maketitle

%%%%%%%%%%%%%%%%%%%%%%%%%%%%%%%%%%%%%%%%%%%%%%%%%%%%%%%%%%%%%%%%%%%%%%%%%%%%%%%%%%%%%%%%%%%%%%%%%%%%%%%%%%%%%%%%%%%%%%%%%%%%%%%%%%%%%%%%%%
\section{Introduction}\label{sec:I}
%%%%%%%%%%%%%%%%%%%%%%%%%%%%%%%%%%%%%%%%%%%%%%%%%%%%%%%%%%%%%%%%%%%%%%%%%%%%%%%%%%%%%%%%%%%%%%%%%%%%%%%%%%%%%%%%%%%%%%%%%%%%%%%%%%%%%%%%%%

The ground state of a spin system and its energy represent valuable information, e.~g., about its low temperature
behaviour. Most research approaches deal with quantum systems, but also the classical limit has found some interest and applications,
see, e.~g., \cite{AL03} - \cite{Setal20}. For classical Heisenberg systems, including Hamiltonians with a Zeeman term due to an external
magnetic field, a rigorous theory has been recently established \cite{SL03} - \cite{SF20} that yields, in principle, all ground states.
However, two restrictions must be made: (1) the dimension $m$ of the ground states found by the theory is {\em per se} not
confined to the physical case of $m\le 3$, and (2) analytical solutions will only be possible for special couplings or small
numbers $N$ of spins. A first application of this theory to frustrated systems with wheel geometry has been given in \cite{FM19} and \cite{FKM19}.

The purpose of the present paper is to give a concise review of the central results of \cite{SL03} -  \cite{SF20}
and to apply the methods outlined there to describe the magnetic behaviour of a spin system
subject to a magnetic field close to the saturation point.
For each field larger than the saturation field $B_{sat}$ all spins will point into the direction
of the field (or opposite the direction, depending on the sign of the Zeeman term), but for values of $B$ slightly
below  $B_{sat}$ the spins will form an ``umbrella" with infinitesimal spread, see, e.~g., Figure \ref{FIGGS4}. It is an obvious goal to calculate
that umbrella in lowest order w.~r.~t.~some sensible expansion parameter $t$.
Another physically interesting property in this connection would be the {\em saturation susceptibility} $\chi_0$,
that is the limit of the susceptibility for $B\uparrow B_{sat}$. Note that numerical calculations close to the saturation point are difficult and do not
yield precise estimates for the spin system's behaviour in lowest order.

In order to investigate the reaction of the spin system to magnetic fields near the saturation point, some case distinctions prove to be necessary.
According to the general theory outlined in \cite{SL03} - \cite{SF20} the various ground states can be obtained
by means of linear combinations of eigenvectors of a so-called {\em dressed $J$-matrix} corresponding to its minimal
eigenvalue. The first case distinction refers to whether the ground state at the saturation point is essentially unique
(non-parabolic case) or not (parabolic case). In the parabolic case the minimal energy $E$ will be a quadratic function
of the magnetization $M$ (hence the name) and consequently the susceptibility will be constant for a certain domain.
In the non-parabolic case the magnetic behaviour in the vicinity of the saturation point can be calculated by means
of a perturbation series up to order four in the parameter $t$ proportional to the spread of the infinitesimal spin umbrella.
This series expansion is easier for coplanar states than for three-dimensional ones, hence the second case distinction.
The form of the infinitesimal spin umbrella close to the saturation point depends on the eigenvectors of the dressed $J$-matrix
in a way to be made more precise below. In the coplanar case there is only one eigenvector that determines the spin umbrella
up to a proportionality factor that can be determined in a straight forward manner.
However, in the three-dimensional case there are two orthogonal eigenvectors and the proportionality
factor has to be replaced by a $2\times 2$-matrix that can only be determined by solving a non-linear system of equations.
These remarks may suffice to illustrate the difference between the coplanar and the three-dimensional case at this point.

After recapitulating, in Section \ref{sec:GT}, the general theory including the aspects relevant for the present problem,
we will, in Section \ref{sec:S}, explain in more details the above-sketched alternative between parabolic and non-parabolic systems
and treat the first ones in Section \ref{sec:P}. After some preliminaries the series expansion for the
non-parabolic case is presented for coplanar ground states, Section \ref{sec:CGS}, and three-dimensional ground states, Section \ref{sec:TGS}.
In both cases, the final equation for saturation susceptibility can be put into a relatively simple common form.
In Section \ref{sec:E} we will present four examples. The first one in Section \ref{sec:IT} is a parabolic irregular tetrahedron
that interestingly deviates from the parabolic behaviour for values of the magnetization $M$ from the interval $0<M_0\le M < M_1<N=4$.
The next three examples are non-parabolic ones.
The isosceles triangle, Section \ref{sec:IT}, has coplanar ground states for all values of $B$ that can be analytically calculated
and hence directly compared with the corresponding perturbation series results. The almost regular cube, Section \ref{sec:EC},
also considered in \cite{SF20} for other reasons, has coplanar ground states close to the saturation field.
Its saturation susceptibility can be determined as the root of a third order equation
and checked with numerical results. Finally, in Section \ref{sec:IH}, we present an irregular octahedron ($N=6$) that is
non-parabolic and admits three-dimensional ground states.
We close with a Summary and Outlook in Section \ref{sec:SO}.

%%%%%%%%%%%%%%%%%%%%%%%%%%%%%%%%%%%%%%%%%%%%%%%%%%%%%%%%%%%%%%%%%%%%%%%%%%%%%%%%%%%%%%%%%%%%%%%%%%%%%%%%%%%%%%%%%%%%%%%%%%%%%%%%%%%%%%%%%%
\section{General theory}\label{sec:GT}
%%%%%%%%%%%%%%%%%%%%%%%%%%%%%%%%%%%%%%%%%%%%%%%%%%%%%%%%%%%%%%%%%%%%%%%%%%%%%%%%%%%%%%%%%%%%%%%%%%%%%%%%%%%%%%%%%%%%%%%%%%%%%%%%%%%%%%%%%%

We will shortly recapitulate the essential results of \cite{S17a}-\cite{S17d} in a form adapted to the present purposes.

%%%%%%%%%%%%%%%%%%%%%%%%%%%%%%%%%%%%%%%%%%%%%%%%%%%%%%%%%%%%%%%%%%%%%%%%%%%%%%%%%%%%%%%%%%%%%%%%%%%%%%%%%%%%%%%%%%%%%%%%%%%%%%%%%%%%%%%%%%
\subsection{Pure Heisenberg systems}\label{sec:HS}
%%%%%%%%%%%%%%%%%%%%%%%%%%%%%%%%%%%%%%%%%%%%%%%%%%%%%%%%%%%%%%%%%%%%%%%%%%%%%%%%%%%%%%%%%%%%%%%%%%%%%%%%%%%%%%%%%%%%%%%%%%%%%%%%%%%%%%%%%%

Let ${\mathbf s}_\mu,\; \mu=1,\ldots,N,$ denote $N$ classical spin vectors of unit length, written as the rows of
an $N\times m$-matrix ${\mathbf s}$ where $m=1,2,3$ is the dimension of the spin vectors. The energy of this system
will be written in the form
\begin{equation}\label{T1}
 H({\mathbf s})=\frac{1}{2}\sum_{\mu,\nu=1}^{N} J_{\mu\nu} {\mathbf s}_\mu\cdot {\mathbf s}_\nu,
\end{equation}
where the $J_{\mu\nu}$ are the entries of a symmetric, real $N\times N$-matrix ${\mathbbm J}$ with vanishing diagonal
elements. In contrast to \cite{S17a}-\cite{S17d} the factor $\frac{1}{2}$ is introduced for convenience.
A {\em ground state} is a spin configuration ${\mathbf s}$ minimizing the energy $ H({\mathbf s})$. If we fix
all vectors ${\mathbf s}_\nu$ of a ground state except a particular one ${\mathbf s}_\mu$, the latter has to minimize
the term
\begin{equation}\label{T2}
 H_\mu\equiv {\mathbf s}_\mu\cdot\left( \sum_{\nu=1}^N J_{\mu\nu} {\mathbf s}_\nu\right)
 \;.
\end{equation}
Hence ${\mathbf s}_\mu$ must be a unit vector opposite to the bracket in (\ref{T2}) and thus has to satisfy
\begin{equation}\label{T3}
  -\kappa_\mu\,{\mathbf s}_\mu=\sum_{\nu=1}^N J_{\mu\nu} {\mathbf s}_\nu
  \;,
\end{equation}
with Lagrange parameters $\kappa_\mu\ge 0$. Upon defining
\begin{equation}\label{T4}
  \overline{\kappa}\equiv\frac{1}{N}\sum_{\mu=1}^{N}\kappa_\mu, \quad \mbox{and } \lambda_\mu\equiv\kappa_\mu-\overline{\kappa}
  \;,
\end{equation}
such that
\begin{equation}\label{T4a}
 \sum_{\mu=1}^{N}\lambda_\mu=0
  \;,
\end{equation}
we may rewrite (\ref{T3}) in the form of an eigenvalue equation
\begin{equation}\label{T5}
  \sum_{\nu=1}^{N}{\mathbbm J}_{\mu\nu}({\boldsymbol\lambda})\,{\mathbf s}_\nu\equiv
   \sum_{\nu=1}^{N}\left( J_{\mu\nu}+\delta_{\mu\nu}\lambda_\nu\right){\mathbf s}_\nu=
   -\overline{\kappa}\,{\mathbf s}_\mu
   \;.
\end{equation}
Here we have introduced the {\em dressed $J$-matrix} ${\mathbbm J}({\boldsymbol\lambda})$ with vanishing trace considered as a function
of the vector ${\boldsymbol\lambda}=(\lambda_1,\ldots,\lambda_N)$ of ``gauge parameters".

We denote by $j_{min}({\boldsymbol\lambda})$ the lowest eigenvalues of ${\mathbbm J}({\boldsymbol\lambda})$ and by
${\mathcal W}_{min}({\boldsymbol\lambda})$ the corresponding eigenspace. It can be shown \cite{S17a} that
the graph of the function
$j_{min}({\boldsymbol\lambda})$, the ``eigenvalue variety",
has a maximum, denoted by $\hat{\jmath}$, that is assumed at a uniquely determined point $\hat{\boldsymbol\lambda}$ such that
\begin{equation}\label{T6}
 E_{min}=\frac{1}{2}\,N \, \hat{\jmath}
\end{equation}
is the ground state energy and that the ground state configuration ${\mathbf s}$ can be obtained as a linear combination
of the corresponding eigenvectors of ${\mathbbm J}(\hat{\boldsymbol\lambda})$. Strictly speaking, the latter
statement has to be restricted to the case where the dimension of ${\mathcal W}_{min}(\hat{\boldsymbol\lambda})$ is
less or equal three, which will be satisfied for all examples considered in this paper.
In the case of one-dimensional ${\mathcal W}_{min}(\hat{\boldsymbol\lambda})$ (collinear ground state) we have
a smooth maximum of  $j_{min}({\boldsymbol\lambda})$, whereas in the cases of a
two- or higher-dimensional ${\mathcal W}_{min}(\hat{\boldsymbol\lambda})$ we have a singular maximum
with a conical structure of $j_{min}({\boldsymbol\lambda})$, at least for some directions in the ${\boldsymbol\lambda}$-space.

Besides the ``ground state gauge"  ${\mathbbm J}(\hat{\boldsymbol\lambda})$ there will be another gauge of the $J$-matrix that will be used,
namely the ``homogeneous gauge" denoted be $J^{(h)}$. It is obtained by subtraction of the corresponding row sums from the diagonal
elements and final addition of the mean row sum $j$:
\begin{equation}\label{T6a}
  J^{(h)}_{\mu\nu}\equiv J_{\mu\nu}+\left( j-\sum_{\lambda}J_{\mu\lambda}\right)\delta_{\mu\nu}
  \;,
\end{equation}
where
\begin{equation}\label{T6b}
 j\equiv \frac{1}{N}\sum_{\mu\nu}J_{\mu\nu}
 \;.
\end{equation}
It follows that $j$ will be an eigenvalue of $J^{(h)}$ corresponding to the eigenvector ${\mathbf 1}=(1,1,\ldots,1)^\top$.
For later use let $j_{min}^{(h)}$ denote the minimal eigenvalue of $J^{(h)}$.

According to the above remarks the ground state configuration ${\mathbf s}$ can be written in the form
\begin{equation}\label{T7}
  {\mathbf s}= W\,\Gamma
  \;,
\end{equation}
where $W$ is an $N\times m$-matrix the columns of which span ${\mathcal W}_{min}(\hat{\boldsymbol\lambda})$,
and $\Gamma$ is a real $m\times m$-matrix. For the $N\times N$ {\em Gram matrix}
\begin{equation}\label{T8}
  G\equiv{\mathbf s}\,{\mathbf s}^\top
\end{equation}
we obtain the following representation:
\begin{equation}\label{T9}
 G\stackrel{(\ref{T7},\ref{T8})}{=}\left(  W\,\Gamma\right)\,\left(  W\,\Gamma\right)^\top=W\,\Gamma\,\Gamma^\top\,W^\top\equiv W\,\Delta\,W^\top
\;.
\end{equation}
Here $\Delta=\Gamma\,\Gamma^\top$ is a positive semi-definite real $m\times m$-matrix that can be obtained
as a solution of the inhomogenous system of linear equations
\begin{equation}\label{T10}
  1={\mathbf s}_\mu\cdot{\mathbf s}_\mu=G_{\mu\mu}\stackrel{(\ref{T9})}{=}\left(W\,\Delta\,W^\top\right)_{\mu\mu},\quad \mu=1,\ldots,m
  \;,
\end{equation}
called ``additionally degeneracy equation" (ADE) in \cite{S17a}.

Let $\Gamma=\sqrt{\Delta}\,R$ be the polar decomposition of $\Gamma$ with $R\in O(m)$, then (\ref{T7}) assumes the form
\begin{equation}\label{T11}
  {\mathbf s}=W\,\sqrt{\Delta}\,R
  \;.
\end{equation}
The rotational/reflectional matrix $R$ in (\ref{T11}) can be chosen quite generally
due to the invariance of $ H({\mathbf s})$ under rotations/reflections. If for each pair of
ground states $({\mathbf s},{\mathbf s}')$ there exists an $R\in O(m)$ such that
${\mathbf s}'={\mathbf s}\,R$ then ${\mathbf s}$  will be called {\em essentially unique}.

%%%%%%%%%%%%%%%%%%%%%%%%%%%%%%%%%%%%%%%%%%%%%%%%%%%%%%%%%%%%%%%%%%%%%%%%%%%%%%%%%%%%%%%%%%%%%%%%%%%%%%%%%%%%%%%%%%%%%%%%%%%%%%%%%%%%%%%%%%
\subsection{Heisenberg-Zeeman systems}\label{sec:HZS}
%%%%%%%%%%%%%%%%%%%%%%%%%%%%%%%%%%%%%%%%%%%%%%%%%%%%%%%%%%%%%%%%%%%%%%%%%%%%%%%%%%%%%%%%%%%%%%%%%%%%%%%%%%%%%%%%%%%%%%%%%%%%%%%%%%%%%%%%%%

In the case of a magnetic field ${\mathbf B}$ that leads to an additional Zeeman term $-{\mathbf B}\cdot {\mathbf S}$
(the sign is chosen as negative without loss of generality)
in the Hamiltonian the ground state problem can be reduced to that of a spin system with a pure Heisenberg Hamiltonian, see \cite{S17d}.
In the first step it is shown that the ground states of the Heisenberg-Zeeman system are among the {\em relative ground states}
of the pure Heisenberg system. These are defined as the ground states under the constraint $\|{\mathbf S}\|^2=M^2$.
The minimal energy $E(M)$ can be extended to an even function defined for $-N\le M\le N$ and, in the smooth case,
the corresponding magnetic field can be obtained as $B(M)=\frac{\partial E}{\partial M}$. The maximal magnetization $M=N$
thus corresponds to the {\em saturation field} $B_{sat}=B(N)$. It can be shown that
\begin{equation}\label{HZS1}
B_{sat}=j-j_{min}^{(h)}
\;,
\end{equation}
see  eq.~(164) in \cite{S17d}, where the missing factor $2$ is due to our modified definition of the energy (\ref{T1}).
Recall that spin systems satisfying $j>j_{min}^{(h)}$ and hence $B_{sat}>0$ have been called ``anti-ferromagnetic" (AF)
in \cite{S17d}. For the present paper this will be generally assumed. Further we may assume  $N\ge 3$, since the case $N=2$ is
completely understood.

In the next step it can be shown that the relative ground states are among the absolute ground states of the pure
Heisenberg system if an auxiliary uniform coupling of strength $\gamma$ is added that leads to a Hamiltonian $H(\gamma)$.
Especially, the phenomenon of saturation can be recovered by varying the uniform coupling. There exists a certain value
$\gamma_0<0$ called the ``critical uniform coupling" such that the following holds:
For $\gamma\le \gamma_0$ the ground state of the system with Hamiltonian $H(\gamma)$ will be
the ferromagnetic ground state corresponding to the eigenvector ${\mathbf 1}$ of the homogeneously gauged $J$-matrix
$J^{(h)}(\gamma)$ and that for $\gamma > \gamma_0$ the ground state will be different from the ferromagnetic one.

In general, the relation between $\gamma$ and $B$ can be complicated. For example, it may happen that the ADE (\ref{T10}) for $H(\gamma)$ has
an $d$-dimensional convex set of solutions such that the corresponding ground states have different magnetization $M$ and different
energy $E(M)$, calculated without uniform coupling. In this way a single value of $\gamma$ may correspond to a whole family
of ground states of the  corresponding Heisenberg-Zeeman system. This will happen in the {\em parabolic case} considered
in Section \ref{sec:P} and $\gamma=\gamma_0$.

On the other hand, it is possible that the ADE (\ref{T10}) for $H(\gamma)$ has only one solution for a certain interval $\gamma_0<\gamma < \gamma_1$
and that there will be a $1:1$-correspondence between uniform coupling strength $\gamma$ and magnetic field $B$ for this interval.
This will happen for the non-parabolic case, see Section \ref{sec:NP}.

In both cases there holds a simple relation between
the saturation field $B_{sat}$ and the critical uniform coupling $\gamma_0$, namely
\begin{equation}\label{HZS2}
 \gamma_0=-\frac{B_{sat}}{N}
 \;,
\end{equation}
following from (\ref{S12a}) and (\ref{HZS1}).

%%%%%%%%%%%%%%%%%%%%%%%%%%%%%%%%%%%%%%%%%%%%%%%%%%%%%%%%%%%%%%%%%%%%%%%%%%%%%%%%%%%%%%%%%%%%%%%%%%%%%%%%%%%%%%%%%%%%%%%%%%%%%%%%%%%%%%%%%%
\section{The saturation alternative}\label{sec:S}
%%%%%%%%%%%%%%%%%%%%%%%%%%%%%%%%%%%%%%%%%%%%%%%%%%%%%%%%%%%%%%%%%%%%%%%%%%%%%%%%%%%%%%%%%%%%%%%%%%%%%%%%%%%%%%%%%%%%%%%%%%%%%%%%%%%%%%%%%%

As explained in Section \ref{sec:HZS} the ground states in the presence of a magnetic field ${\mathbf B}$ are among the
ground states assumed by the pure Heisenberg spin system with an auxiliary uniform coupling of strength $\gamma$.
If $\gamma $ is negative and arbitrarily large in absolute value all spins will be aligned into the direction of the field
and the maximal magnetization $M=N$ is reached. Let $\gamma_0$ be the maximal value where this happens such that for
$\gamma> \gamma_0$ the ground state will not be fully aligned and $M<N$.
The corresponding critical field is called the {\em saturation field} $B_{sat}$, see (\ref{HZS1})
and  (\ref{S7}), (\ref{SS2}) and (\ref{TGSSS2}) below.

We consider a matrix ${\mathbbm J}$ of coupling coefficients that depends on the gauge parameters ${\boldsymbol{\lambda}}$
and an auxiliary uniform coupling coefficient $\gamma$. This dependence will be written as
\begin{equation}\label{S1}
  {\mathbbm J}_{\mu\nu}({\boldsymbol{\lambda}},\gamma)= {\mathbbm J}_{\mu\nu}({\mathbf 0},0)+\delta_{\mu\nu}\,\lambda_\nu+\gamma\,\Xi_{\mu\nu}
\;\quad \mu,\nu=1,\ldots,N
\;,
\end{equation}
where
\begin{equation}\label{S2}
  \Xi_{\mu\nu}\equiv
1-\delta_{\mu\nu}=
\left\{
 \begin{array}{r@{\quad \mbox{if} \quad}l}
1 & \mu\neq\nu,\\
0& \mu=\nu.
 \end{array}
 \right.
\end{equation}
It follows from the above remarks that for $\gamma<\gamma_0$ the vector ${\mathbf 1}\equiv (1,1,\ldots,1)^\top$ will be the ground state
of the spin system characterized by (\ref{S1}) and the corresponding ground state gauge will be given by
\begin{equation}\label{S3}
\kappa_\mu^{(0)}=-\sum_\nu  {\mathbbm J}_{\mu\nu}({\mathbf 0},0),\quad
\lambda_\mu^{(0)}
=\kappa_\mu^{(0)}-\frac{1}{N}\sum_\nu \kappa_\nu^{(0)}\equiv \kappa_\mu^{(0)}+j
\;,
\end{equation}
for $\mu=1,\ldots,N$ and  $j$ denoting the mean row sum of $ {\mathbbm J}_{\mu\nu}({\mathbf 0},0)$.
It follows that ${\mathbbm J}_{\mu\nu}({\boldsymbol\lambda}^{(0)},0)$ is homogeneously gauged, i.~e.,
\begin{equation}\label{S4}
 {\mathbbm J}({\boldsymbol\lambda}^{(0)},0)=J^{(h)}
 \;,
\end{equation}
and hence
\begin{equation}\label{S5}
  {\mathbbm J}({\boldsymbol\lambda}^{(0)},0)\,{\mathbf 1}=j\,{\mathbf 1}
  \;.
\end{equation}
By definition the matrix $J^{(h)}$ has constant row (column) sums and hence commutes with $\Xi$.
Since $\Xi$ has also constant row sums, equal to $N-1$, it follows that ${\mathbbm J}({\boldsymbol{\lambda}^{(0)}},\gamma)$
is also homogeneously gauged and satisfies
\begin{equation}\label{S6}
 {\mathbbm J}({\boldsymbol{\lambda}^{(0)}},\gamma)\,{\mathbf 1}=(j+(N-1)\gamma)\,{\mathbf 1}\equiv j(\gamma)\,{\mathbf 1}
 \;.
\end{equation}

\begin{figure}[t]
\centering
\includegraphics[width=0.70\linewidth]{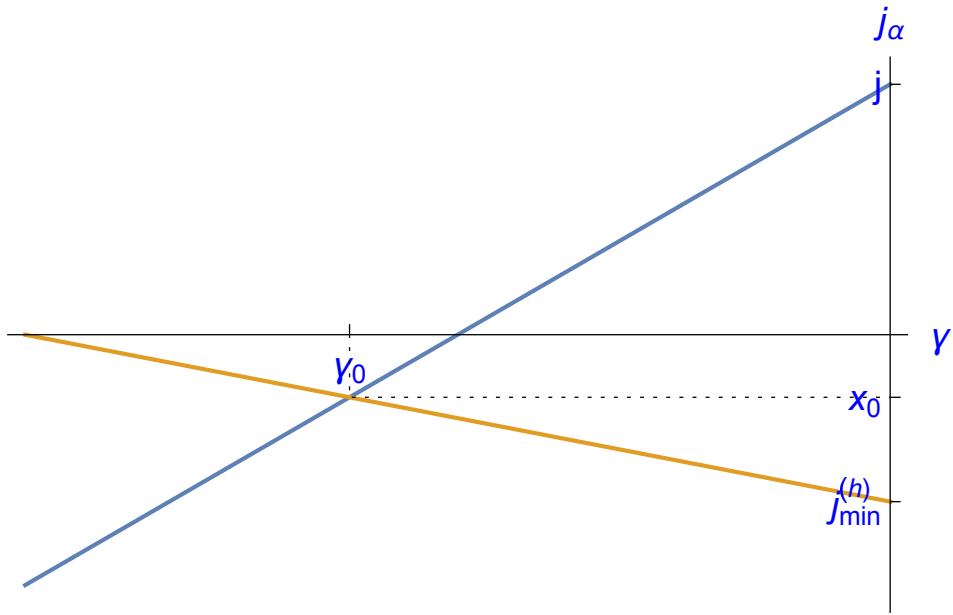}
\caption{Schematic representation of the linear $\gamma$-dependence of the two eigenvalues
$j(\gamma)=j+(N-1)\,\gamma$ (blue line) and $j_0(\gamma)=j_{min}^{(h)}-\gamma$ (dark yellow line).
The two lines meet at $\gamma=\gamma_0$ given by  (\ref{S12b}) thereby defining the critical value
of the saturation point.
}
\label{FIGEV}
\end{figure}

For sufficiently large negative $\gamma$ the eigenvalue $j(\gamma)$ will be the lowest eigenvalue of
${\mathbbm J}({\boldsymbol{\lambda}^{(0)}},\gamma)$ and ${\mathbf 1}$ will be the ground state. This property is
lost if another eigenvalue assumes the role of the lowest one. Hence the critical value $\gamma_0$ can be characterized
as the lowest value of $\gamma$ such that $j(\gamma_0)$ becomes degenerate. To determine $\gamma_0$ let us consider the (possibly degenerate)
lowest eigenvalue $j_{min}^{(h)}$ of $J^{(h)}= {\mathbbm J}({\boldsymbol\lambda}^{(0)},0)$ and an arbitrary normalized
corresponding eigenvector ${\boldsymbol\xi}$:
\begin{equation}\label{S7}
J^{(h)}\,{\boldsymbol\xi}={\mathbbm J}({\boldsymbol\lambda}^{(0)},0)\,{\boldsymbol\xi}=j_{min}^{(h)}\, {\boldsymbol\xi}
\;.
\end{equation}
Due to the general assumption  $j>j_{min}^{(h)}$ we conclude that ${\boldsymbol\xi}\perp{\mathbf 1}$, i.~e.,
\begin{equation}\label{S8}
  \sum_\mu \xi_\mu=0
  \;.
\end{equation}
Due to
\begin{equation}\label{S9}
 \Xi = |{\mathbf 1}\rangle\langle {\mathbf 1}|-{\mathbbm 1}
 \;,
\end{equation}
${\boldsymbol\xi}$ will also be an eigenvector of $\Xi$ with eigenvalue $-1$:
\begin{equation}\label{S10}
 \Xi\,{\boldsymbol\xi}= |{\mathbf 1}\rangle\underbrace{\langle {\mathbf 1}|{\boldsymbol\xi}\rangle}_0 -{\boldsymbol\xi}=-{\boldsymbol\xi}
 \;.
 \end{equation}
Hence
\begin{equation}\label{S11}
{\mathbbm J}({\boldsymbol\lambda}^{(0)},\gamma)\,{\boldsymbol\xi}=
{\mathbbm J}({\boldsymbol\lambda}^{(0)},0)\,{\boldsymbol\xi}+\gamma\,\Xi\,{\boldsymbol\xi}
\stackrel{(\ref{S7},\ref{S10})}{=}
\left(j_{min}^{(h)}-\gamma\right)\,{\boldsymbol\xi}\equiv j_0(\gamma)\,{\boldsymbol\xi}
\;.
\end{equation}
This means that, for $\gamma<0$, the two eigenvalues $j(\gamma)$ and $j_0(\gamma)$ of ${\mathbbm J}({\boldsymbol\lambda}^{(0)},\gamma)$
behave differently, the first one decreases with growing $|\gamma|$ and the second one increases, see Figure \ref{FIGEV}.
The two lines in Figure \ref{FIGEV} representing $j(\gamma)$ and $j_0(\gamma)$ intersect at the critical value $\gamma_0$ defined by $j(\gamma_0)=j_0(\gamma_0)$ according to
\begin{eqnarray}\label{S12a}
 && j+(N-1)\,\gamma_0= j_{min}^{(h)}-\gamma_0\equiv x_0\\
 \label{S12b}
  &\Leftrightarrow& \gamma_0=\frac{1}{N}\left(j_{min}^{(h)}-j \right)<0
  \;.
\end{eqnarray}

Actually, the $\gamma$-dependence of $j_\alpha(\gamma)=j_\alpha -\gamma$ holds for every eigenvalue $j_\alpha$
of $J^{(h)}$ different from $j$ and leads to corresponding intersections with $j(\gamma)$ at $\gamma_\alpha=\frac{1}{N}\left(j_\alpha-j \right)$.
The critical value $\gamma_0$ will be given by the lowest one of these $\gamma_\alpha$ and hence by the lowest eigenvalue
$j_0=j_{min}^{(h)}$ of $J^{(h)}$. Moreover, the value $x_0$ in (\ref{S12a}) will be the lowest eigenvalue of
${\mathbbm J}_{\mu\nu}({\boldsymbol{\lambda}^{(0)}},\gamma)$ for $\gamma\le \gamma_0$.

To summarize: For $\gamma\le \gamma_0$ the state ${\mathbf 1}$ will be the ground state of the spin system characterized by the dressed $J$-matrix
${\mathbbm J}_{\mu\nu}({\boldsymbol{\lambda}^{(0)}},\gamma)$ and hence all spins
are aligned parallel to the magnetic field. For $\gamma>\gamma_0$ this is no longer the case and hence $\gamma_0$
is the critical uniform coupling defining what we will call the saturation point.

Further, the following alternative occurs: Either at $\gamma=\gamma_0$ the ground state ${\mathbf 1}$ is essentially unique,
i.~e., the corresponding ADE (\ref{T10}) has exactly one solution, or, there exists at least one other ground state at $\gamma=\gamma_0$
and hence the convex set ${\mathcal S}_{ADE}$ of solutions of (\ref{T10}) contains more than one, and hence infinitely many points.
We conjecture that this alternative is identical to the distinction between ``continuous reduction" and ``discontinuous reduction"
made in \cite{S17d}.

In the first case we have a smooth family ${\mathbf s}(\gamma)$ of unique ground states
for some interval $\gamma_0<\gamma<\gamma_0+\varepsilon$ satisfying ${\mathbf s}(\gamma_0)={\mathbf 1}$
and may investigate the magnetic behaviour of the spin system
in the vicinity of the saturation point by means of a perturbational series, see Section \ref{sec:NP}.
The susceptibility at the saturation point assumes the form
\begin{equation}\label{S13}
  \chi_0=\frac{N}{B_{sat}+N\,k^2}
\;,
\end{equation}
see (\ref{SS4d}), (\ref{TGSSS4d}) and Figure \ref{FIGBM1}.

In the second case
we have another smooth family ${\mathbf s}(t)$ of ground states such that ${\mathbf s}(0)={\mathbf 1}$ but this family can be
constructed solely from states given by  ${\mathcal S}_{ADE}$ at $\gamma=\gamma_0$, see Section \ref{sec:P}.
The family ${\mathbf s}(t)$ may include the absolute ground state or not.
Moreover, for this family of ground states the energy (without uniform coupling) will be a
simple quadratic function $E(M)$ of the magnetization $M$, a property that has been called
``parabolicity" in \cite{S17d}. Consequently, near the saturation point the susceptibility will be constant
assuming the value
\begin{equation}\label{S14}
  \chi= \frac{N}{B_{sat}}
\;,
\end{equation}
see (\ref{S8}).

It is not clear whether the above ``saturation alternative" covers all possibilities. In the parabolic case
it may happen that the family ${\mathbf s}(t)$ contains un-physical ground states of dimension greater than three, and
that the physical ground states do not give rise to a quadratic function $E(M)$. The AF icosahedron is an example, see \cite{SSSL05}.

%%%%%%%%%%%%%%%%%%%%%%%%%%%%%%%%%%%%%%%%%%%%%%%%%%%%%%%%%%%%%%%%%%%%%%%%%%%%%%%%%%%%%%%%%%%%%%%%%%%%%%%%%%%%%%%%%%%%%%%%%%%%%%%%%%%%%%%%%%
\section{Parabolic case}\label{sec:P}
%%%%%%%%%%%%%%%%%%%%%%%%%%%%%%%%%%%%%%%%%%%%%%%%%%%%%%%%%%%%%%%%%%%%%%%%%%%%%%%%%%%%%%%%%%%%%%%%%%%%%%%%%%%%%%%%%%%%%%%%%%%%%%%%%%%%%%%%%%

According to Section \ref{sec:S}, at the saturation point the dressed $J$-matrix
${\mathbbm J}({\boldsymbol\lambda}^{(0)},\gamma_0)$ has a degenerate minimal eigenvalue
$x_0$ and a corresponding eigenspace ${\mathcal E}_0\equiv {\mathcal W}_{min}({\boldsymbol\lambda}^{(0)})$
containing the vector ${\mathbf 1}$ that represents the ferromagnetic ground state.
We now consider the case where it is possible to obtain another $m$-dimensional ground state ${\boldsymbol\sigma}$ by means of
linear combinations of vectors of ${\mathcal E}_0$.
Recall from the general theory that these linear combinations are encoded
in some positively semi-definite $m\times m$-matrix $\Delta$ that solves the ADE (\ref{T10}). We hence consider the case where
the compact convex solution set ${\mathcal S}_{ADE}$ of (\ref{T10}) contains more than one point.

Since the vectors $\left( {\boldsymbol\sigma}_\mu^{(i)}\right)_{\mu=1,\ldots,N}$ lie in ${\mathcal E}_0$ for $i=1,\ldots,m,$ we have
\begin{equation}\label{P0}
 \sum_\nu {\mathbbm J}_{\mu\nu}({\boldsymbol\lambda}^{(0)},\gamma_0)\,{\boldsymbol\sigma}_\nu^{(i)}=x_0\,{\boldsymbol\sigma}_\mu^{(i)}
 \;,
\end{equation}
for $i=1,\ldots,m,$ and $\mu=1,\ldots,N$.

We define a family of $(m+1)$-dimensional ground states ${\mathbf s}(t)$ that interpolates between ${\mathbf 1}$ and ${\boldsymbol\sigma}$:
\begin{equation}\label{P1}
  {\mathbf s}_\mu(t) ={\sqrt{1-t^2} \choose  {\boldsymbol\sigma}_\mu\,t },\quad 0\le t \le 1,\; \mbox{and } \mu=1,\ldots,N
  \;.
\end{equation}
It is clear that the $ {\mathbf s}_\mu(t)$ are unit vectors. The total spin is obtained as
\begin{equation}\label{P2}
  {\mathbf S}(t)=\sum_\mu {\mathbf s}_\mu(t) ={ N\,\sqrt{1-t^2}\choose \sum_\mu {\boldsymbol\sigma}_\mu\,t }
  \equiv { N\,\sqrt{1-t^2}\choose  {\boldsymbol\Sigma}\,t }
  \;,
\end{equation}
and yields the squared magnetization
\begin{equation}\label{P3}
  M(t)^2={\mathbf S}(t)\cdot{\mathbf S}(t)=N^2\,(1-t^2)+ \Sigma^2\,t^2
  \;.
\end{equation}
Using
\begin{equation}\label{P4}
  {\mathbf s}_\mu(t)\cdot {\mathbf s}_\nu(t) =1-t^2+{\boldsymbol\sigma}_\mu\cdot{\boldsymbol\sigma}_\nu\,t^2
\end{equation}
for all $\mu,\nu=1,\ldots,N$
we calculate the energy (without the uniform coupling):
\begin{eqnarray}
\label{P5a}
  E(t) &=& \frac{1}{2}\sum_{\mu\nu}J_{\mu\nu}^{(h)} {\mathbf s}_\mu(t)\cdot {\mathbf s}_\nu(t) \\
  \label{P5b}
   &\stackrel{(\ref{P4})}{=}& \frac{1}{2} \left(
   \sum_{\mu\nu}J_{\mu\nu}^{(h)}  (1-t^2)+\sum_{\mu\nu}J_{\mu\nu}^{(h)} {\boldsymbol\sigma}_\mu\cdot{\boldsymbol\sigma}_\nu\,t^2
   \right)\\
  \label{P5c}
   &\stackrel{(\ref{S4},\ref{S5})}{=}& \frac{1}{2} N j (1-t^2)+\frac{1}{2}\left[
   \sum_{\mu\nu}\left(    J_{\mu\nu}^{(h)} +\gamma_0\Xi_{\mu\nu}\right){\boldsymbol\sigma}_\mu\cdot{\boldsymbol\sigma}_\nu-
    \gamma_0\sum_{\mu\nu}\Xi_{\mu\nu} {\boldsymbol\sigma}_\mu\cdot{\boldsymbol\sigma}_\nu
   \right]\,t^2\\
   \label{P5d}
   &\stackrel{(\ref{S1},\ref{S2})}{=}&\frac{1}{2} N j (1-t^2)+\frac{1}{2}\left[
   \sum_{\mu\nu}{\mathbbm J}_{\mu\nu}\left({\boldsymbol\lambda}^{(0)},\gamma_0\right){\boldsymbol\sigma}_\mu\cdot{\boldsymbol\sigma}_\nu-
    \gamma_0\sum_{\mu\nu}\left(1-\delta_{\mu\nu} \right) {\boldsymbol\sigma}_\mu\cdot{\boldsymbol\sigma}_\nu
   \right]\,t^2\\
   \label{P5e}
   &\stackrel{(\ref{P0})}{=}&\frac{1}{2} \left( N j (1-t^2)+ N\,x_0\,t^2-\gamma_0\,\Sigma^2\,t^2+\gamma_0\, N\, t^2\right)\\
   &\stackrel{(\ref{S12a})}{=}& \frac{1}{2}\, N\,j+\frac{\gamma_0}{2}\,\left( N^2-\Sigma^2\right)t^2\\
   \label{P5f}
   &\stackrel{(\ref{P3})}{=}&\frac{1}{2}\, N\,j+\frac{\gamma_0}{2}\,\left( N^2-M(t)^2\right)\\
   \label{P5g}
   &\stackrel{(\ref{S12b})}{=}& \frac{1}{2}\, N\,j_{min}^{(h)}+\frac{j-j_{min}^{(h)}}{2\,N}\, M(t)^2
   \;.
\end{eqnarray}
Recall that the a spin system satisfying the last equation has been called ``parabolic"  in \cite{S17d}, eq.~(164).
The missing factor $\frac{1}{2}$ is due to our modified definition of the energy in (\ref{T1}). Another difference
is that in \cite{S17d} the validity of (\ref{P5g}) was required for the interval $\check{\mu}\le M\le N$,
$\check{\mu}$ denoting the magnetization corresponding to the ``threshold field" $B_{thr}$, see \cite{S17d},
whereas we have only proven (\ref{P5g}) for $M(1)\le M(t)\le M(0)$. We will provide an example in Section \ref{sec:IT}
showing that the condition of parabolicity may be only satisfied for a smaller interval than required in \cite{S17d} and hence the definition
of ``parabolicity" should be accordingly weakened.

As an immediate consequence of (\ref{P5g}) we note that for the considered one-parameter family the magnetic field obeys
\begin{equation}\label{P6}
  B(t)=\frac{\partial E}{\partial M}= \frac{j-j_{min}^{(h)}}{N}\, M(t)
  \;,
\end{equation}
which yields the saturation field
\begin{equation}\label{P7}
 B_{sat}=B(0) = \frac{j-j_{min}^{(h)}}{N}\, M(0)= j-j_{min}^{(h)}
\end{equation}
in accordance with (\ref{HZS1}).

For the susceptibility we obtain the constant value
\begin{equation}\label{P8}
\chi=\frac{\partial M}{\partial B}= \frac{N}{j-j_{min}^{(h)}}\stackrel{(\ref{P7})}{=}\frac{N}{B_{sat}}
\stackrel{(\ref{S12b})}{=}-\frac{1}{\gamma_0}
\;.
\end{equation}

%%%%%%%%%%%%%%%%%%%%%%%%%%%%%%%%%%%%%%%%%%%%%%%%%%%%%%%%%%%%%%%%%%%%%%%%%%%%%%%%%%%%%%%%%%%%%%%%%%%%%%%%%%%%%%%%%%%%%%%%%%%%%%%%%%%%%%%%%%
\section{Non-parabolic case}\label{sec:NP}
%%%%%%%%%%%%%%%%%%%%%%%%%%%%%%%%%%%%%%%%%%%%%%%%%%%%%%%%%%%%%%%%%%%%%%%%%%%%%%%%%%%%%%%%%%%%%%%%%%%%%%%%%%%%%%%%%%%%%%%%%%%%%%%%%%%%%%%%%%

\begin{figure}[t]
\centering
\includegraphics[width=0.70\linewidth]{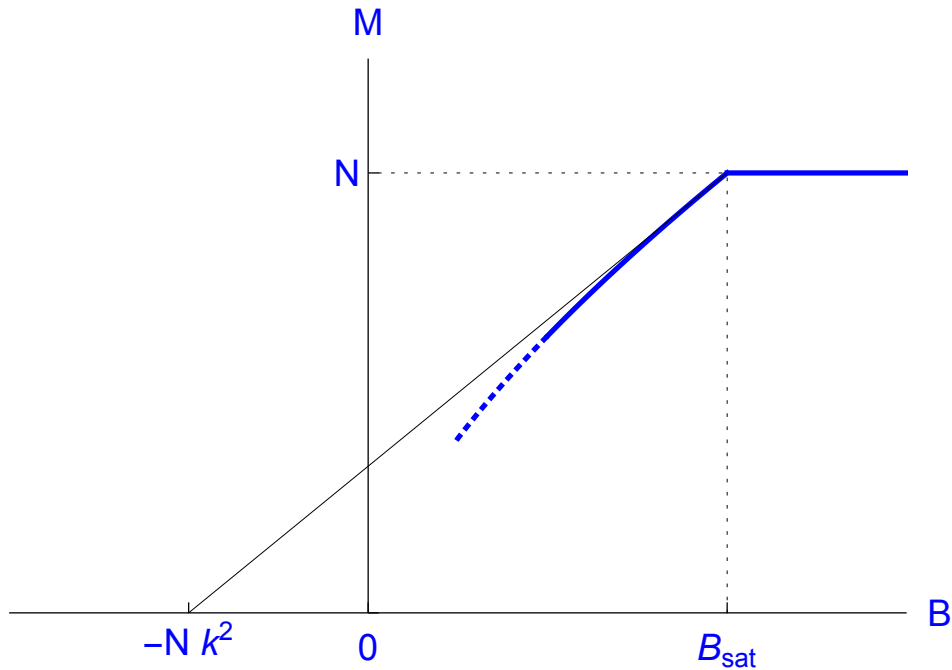}
\caption{Schematic representation of the magnetization $M$ as a function of the magnetic field $B$ in the vicinity
of the saturation point $(B_{sat},N)$ in the non-parabolic case. The saturation susceptibility $\chi_0$ is the slope of the function $M(B)$
at the saturation point. According to (\ref{SS4d}) and  (\ref{TGSSS4d})
it assumes the form  $\chi_0=\frac{N}{B_{sat}+Nk^2}$ and hence the tangent to $M(B)$ at the
saturation point intersects the $B$-axis at $-Nk^2$.
}
\label{FIGBM1}
\end{figure}

According to Section \ref{sec:S}, at the saturation point the dressed $J$-matrix
${\mathbbm J}({\boldsymbol\lambda}^{(0)},\gamma_0)$ has a degenerate eigenvalue
$x_0$ and a corresponding eigenspace ${\mathcal E}_0$ containing the vector ${\mathbf 1}$ that represents the ferromagnetic ground state.
We now consider the case where it is not possible to obtain another $m$-dimensional ground state ${\boldsymbol\sigma}$ by means of
linear combinations of vectors of ${\mathcal E}_0$.
Recall from the general theory that these linear combinations are encoded
in some positively semi-definite $m\times m$-matrix $\Delta$ that solves the ADE (\ref{T10}). We hence consider the case where
the compact convex solution set ${\mathcal S}_{ADE}$ of (\ref{T10}) contains exactly one point.

Generally, we denote the subspace of ${\mathcal E}_0$ orthogonal to ${\mathbf 1}$ by ${\mathcal E}_1$ such that
\begin{equation}\label{NP1}
 {\mathcal E}_0={\mathcal E}_1\oplus {\mathbbm R}\,{\mathbf 1}
 \;.
\end{equation}

In this section we will assume local analyticity, i.~e., that for some interval $\gamma_0<\gamma<\gamma+\varepsilon$
the physically relevant quantities can be expanded into power series w.~r.~t.~a certain parameter $t$.
However, $t$ cannot be chosen as $\gamma-\gamma_0$ but rather as $t=\sqrt{\gamma-\gamma_0}$. This can be made plausible
by the square root in the representation of the ground state as ${\mathbf s}=W\,\sqrt{\Delta}\,R$, see (\ref{T11}). Even if the matrix
$\Delta$ could be expanded into a power series w.~r.~t.~$\gamma-\gamma_0$, the ground state itself can only be represented by
a $t$-series with $t=\sqrt{\gamma-\gamma_0}$. This also explains why we need the fourth order expansion to calculate the
saturation susceptibility $\chi_0$. Due to $B=\frac{\partial E}{\partial M}$ and  $\chi=\frac{\partial M}{\partial B}$
the second order would suffice, but this is the second order of the expansion of $E$ and $M$ w.~r.~t.~the variable $\gamma-\gamma_0=t^2$.
The fact that the ground state varies with $t$ whereas the minimal energy varies with $t^2$ also explains the poor quality of
numerical ground state determination close to the saturation point.

For the critical value $\gamma=\gamma_0$ the vector ${\mathbf 1}\equiv (1,1,\ldots,1)$ will still be an eigenvector of the dressed
J-matrix ${\mathbbm J}({\boldsymbol \lambda}^{(0)},\gamma_0)$. The gauge parameters ${\boldsymbol \lambda}^{(0)}$ and the corresponding
eigenvalue $x_0$ have already been  calculated, see (\ref{S3}) and (\ref{S12a}).

We will make the case distinction according to whether the ground states for $\gamma_0<\gamma<\gamma+\varepsilon$
are two- or three-dimensional. This is sufficient to cover the physical cases but higher-dimensional ground states could be
calculated by analogous methods.

%%%%%%%%%%%%%%%%%%%%%%%%%%%%%%%%%%%%%%%%%%%%%%%%%%%%%%%%%%%%%%%%%%%%%%%%%%%%%%%%%%%%%%%%%%%%%%%%%%%%%%%%%%%%%%%%%%%%%%%%%%%%%%%%%%%%%%%%%%
\subsection{Coplanar ground states}\label{sec:CGS}
%%%%%%%%%%%%%%%%%%%%%%%%%%%%%%%%%%%%%%%%%%%%%%%%%%%%%%%%%%%%%%%%%%%%%%%%%%%%%%%%%%%%%%%%%%%%%%%%%%%%%%%%%%%%%%%%%%%%%%%%%%%%%%%%%%%%%%%%%%

We assume that the eigenspace ${\mathcal E}_0$ of ${\mathbbm J}_{\mu\nu}({\boldsymbol{\lambda}}^{(0)},\gamma_0)$
corresponding to the lowest eigenvalue $x_0$
is two-dimensional and hence the subspace ${\mathcal E}_1$ according to (\ref{NP1}) will be one-dimensional.
Let ${\boldsymbol\xi}$  be a fixed normalized basis vector in ${\mathcal E}_1$.

%%%%%%%%%%%%%%%%%%%%%%%%%%%%%%%%%%%%%%%%%%%%%%%%%%%%%%%%%%%%%%%%%%%%%%%%%%%%%%%%%%%%%%%%%%%%%%%%%%%%%%%%%%%%%%%%%%%%%%%%%%%%%%%%%%%%%%%%%%
\subsubsection{Notations and first results}\label{sec:PN}
%%%%%%%%%%%%%%%%%%%%%%%%%%%%%%%%%%%%%%%%%%%%%%%%%%%%%%%%%%%%%%%%%%%%%%%%%%%%%%%%%%%%%%%%%%%%%%%%%%%%%%%%%%%%%%%%%%%%%%%%%%%%%%%%%%%%%%%%%%

Recall that the $J$-matrix depending on the gauge parameters ${\boldsymbol\lambda}$ and the uniform coupling strength $\gamma$
assumes the form
\begin{equation}\label{PN1}
  {\mathbbm J}_{\mu\nu}({\boldsymbol{\lambda}},\gamma)= {\mathbbm J}_{\mu\nu}({\mathbf 0},0)+\delta_{\mu\nu}\,\lambda_\nu+\gamma\,\Xi_{\mu\nu}
\;\quad \mu,\nu=1,\ldots,N
\;,
\end{equation}
where
\begin{equation}\label{PN2}
  \Xi_{\mu\nu}\equiv
1-\delta_{\mu\nu}=
\left\{
 \begin{array}{r@{\quad \mbox{if} \quad}l}
1 & \mu\neq\nu,\\
0& \mu=\nu.
 \end{array}
 \right.
\end{equation}
We set
\begin{equation}\label{PN3}
J_{\mu\nu}^{(0)} \equiv {\mathbbm J}_{\mu\nu}({\mathbf 0},\gamma_0)
\;,
\end{equation}
and consider the one-parameter families
\begin{eqnarray}\label{PN4a}
 J_{\mu\nu}(t)&=&J_{\mu\nu}^{(0)}+t^2\,\Xi_{\mu\nu}\;,\\
\label{PN4b}
{\mathbf s}_\mu(t)&=& \sum_{n=0,1,2,\ldots}t^n\,{\mathbf s}_\mu^{(n)}\;,\\
\label{PN4c}
&=&{1\choose 0}+t\,{0\choose y_\mu^{(1)}}+t^2\,{ x_\mu^{(2)}\choose 0}+t^3\,{0\choose y_\mu^{(3)}}+t^4\,{ x_\mu^{(4)}\choose 0}+\ldots\;,\\
\label{PN4d}
\kappa_\mu(t)&=& \sum_{n=0,2,4,\ldots}t^n\,{\kappa}_\mu^{(n)}\;,\\
\label{PN4e}
x(t)&=&-\frac{1}{N}\sum_\mu \kappa_\mu(t)= \sum_{n=0,2,4,\ldots}t^n\,x_n
\;,
\end{eqnarray}
for $\mu=1,\ldots,N$. The condition $\|{\mathbf s}_\mu(t)\|=1$ for all $\mu=1,\ldots,N$
entails an infinite number of identities for the $x_\mu^{(n)}$, the first two of which read
\begin{eqnarray}
\label{PN5a}
 x_\mu^{(2)} &=& -\frac{1}{2}\,y_\mu^{(1)2}\;,\\
\label{PN5b}
 x_\mu^{(4)} &=& -\left( y_\mu^{(1)}\ y_\mu^{(3)}+\frac{1}{8}y_\mu^{(1)4}\right)
\;.
\end{eqnarray}
In the ground state configuration the total spin ${\mathbf S}(t)$ will point into the direction ${1\choose 0}$ of the field and hence
\begin{equation}\label{PN5c}
 {\mathbf S}(t) = \sum_\mu {\mathbf s}_\mu(t)\equiv {M(t)\choose 0}
 \;,
\end{equation}
which yields the series representation of the magnetization
\begin{equation}\label{PN5d}
M(t)=N+t^2 M^{(2)}+t^4\,M^{(4)}+\ldots
\stackrel{(\ref{PN5c},\ref{PN4c})}{=}
\sum_{n=0,2,4,\ldots} t^n\;\; \sum_\mu x_\mu^{(n)}
\;.
\end{equation}
We note that Eqs.~(\ref{PN5c}) and (\ref{PN4c}) imply
\begin{equation}\label{PN5e}
 \sum_\mu y_\mu^{(n)} = 0 \quad \mbox{ for all odd } n
 \;.
\end{equation}
Further we consider the energy (without the auxiliary uniform coupling)
\begin{equation}\label{PN5f}
 E(t)=\frac{1}{2}\sum_{\mu\nu} {\mathbbm J}_{\mu\nu}({\mathbf 0},0)\,{\mathbf s}_\mu(t)\cdot {\mathbf s}_\mu(t)
 = E^{(0)}+t^2 E^{(2)}+t^4 E^{(4)}+\ldots
\end{equation}
The  $t$-series for $M(t)$ and $E(t)$ contain only even terms since the scalar product of two terms of different parity in (\ref{PN4b}) vanishes.

%%%%%%%%%%%%%%%%%%%%%%%%%%%%%%%%%%%%%%%%%%%%%%%%%%%%%%%%%%%%%%%%%%%%%%%%%%%%%%%%%%%%%%%%%%%%%%%%%%%%%%%%%%%%%%%%%%%%%%%%%%%%%%%%%%%%%%%%%%
\subsubsection{Perturbation series}\label{sec:PS}
%%%%%%%%%%%%%%%%%%%%%%%%%%%%%%%%%%%%%%%%%%%%%%%%%%%%%%%%%%%%%%%%%%%%%%%%%%%%%%%%%%%%%%%%%%%%%%%%%%%%%%%%%%%%%%%%%%%%%%%%%%%%%%%%%%%%%%%%%%

We rewrite Eq.~(\ref{T3}) in the form
\begin{equation}\label{PN6}
  \sum_\nu J_{\mu\nu}(t)\,{\mathbf s}_\nu(t)=-\kappa_\mu(t)\,{\mathbf s}_\mu(t)
\;,
\end{equation}
expand both sides into powers of $t$ and equate identical powers.
The following subsections are devoted to the evaluation of (\ref{PN6}) for orders $t^0,\ldots,t^4$.
This method is closely analogous to the usual Rayleigh-Schr\"odinger perturbation theory of eigenvalue equations in quantum mechanics.

%%%%%%%%%%%%%%%%%%%%%%%%%%%%%%%%%%%%%%%%%%%%%%%%%%%%%%%%%%%%%%%%%%%%%%%%%%%%%%%%%%%%%%%%%%%%%%%%%%%%%%%%%%%%%%%%%%%%
\subsubsection{Terms $O(t^0)$:}\label{sec:PS0}
%%%%%%%%%%%%%%%%%%%%%%%%%%%%%%%%%%%%%%%%%%%%%%%%%%%%%%%%%%%%%%%%%%%%%%%%%%%%%%%%%%%%%%%%%%%%%%%%%%%%%%%%%%%%%%%%%%%%

By evaluating  (\ref{PN6})  for $t=0$ we recover the results of Section \ref{sec:S} concerning the ground state problem
at the saturation point. Especially,
\begin{equation}\label{PN19}
 E^{(0)}=\frac{1}{2}\sum_{\mu\nu}  {\mathbbm J}_{\mu\nu}({\mathbf 0},0)\,{\mathbf s}_\mu^{(0)}\cdot {\mathbf s}_\mu^{(0)}
 =\frac{1}{2}\sum_{\mu\nu}  {\mathbbm J}_{\mu\nu}({\mathbf 0},0)\,{1\choose 0}\cdot {1\choose 0}
 =\frac{1}{2}\sum_{\mu\nu}  {\mathbbm J}_{\mu\nu}({\mathbf 0},0) \stackrel{(\ref{T6b})}{=}\frac{N\,j}{2}
 \;,
\end{equation}
in accordance with (\ref{T6}).

%%%%%%%%%%%%%%%%%%%%%%%%%%%%%%%%%%%%%%%%%%%%%%%%%%%%%%%%%%%%%%%%%%%%%%%%%%%%%%%%%%%%%%%%%%%%%%%%%%%%%%%%%%%%%%%%%%%%
\subsubsection{Terms $O(t^1)$:}\label{sec:PS1}
%%%%%%%%%%%%%%%%%%%%%%%%%%%%%%%%%%%%%%%%%%%%%%%%%%%%%%%%%%%%%%%%%%%%%%%%%%%%%%%%%%%%%%%%%%%%%%%%%%%%%%%%%%%%%%%%%%%%

The $t$-linear terms of (\ref{PN6}) read:
\begin{equation}\label{PN20}
  \sum_\nu J_{\mu\nu}^{(0)}\,{\mathbf s}_\nu^{(1)}=-\kappa_\mu^{(0)}\,{\mathbf s}_\mu^{(1)}
  \;.
\end{equation}
Using (\ref{PN4c}) this means
\begin{equation}\label{PN21}
  \sum_\nu J_{\mu\nu}^{(0)}\,y_\nu^{(1)}=-\kappa_\mu^{(0)}\,y_\mu^{(1)}
  \;,
\end{equation}
or, due to (\ref{T4}) and (\ref{S12a}),
\begin{equation}\label{PN22}
  \sum_\nu {\mathbbm J}_{\mu\nu}({\boldsymbol\lambda}^{(0)},\gamma_0)\,y_\nu^{(1)}=x_0\,y_\mu^{(1)}
  \;.
\end{equation}
Hence $y^{(1)}$ is an eigenvector of ${\mathbbm J}_{\mu\nu}({\boldsymbol\lambda}^{(0)},\gamma_0)$ corresponding to its lowest eigenvalue $x_0$.
According to (\ref{PN5e}) this eigenvector $y^{(1)}$ is orthogonal to ${\mathbf 1}$ and hence proportional to ${\boldsymbol \xi}$:
\begin{equation}\label{PN23}
  y_\mu^{(1)}=X\,\xi_\mu,\quad \mbox{for some constant } X>0
\;,
\end{equation}
and all $\mu=1,\ldots,N$. $X$ may be chosen positive since ${\boldsymbol\xi}$ is only unique up to a sign.
The value of $X$ will be determined later. For the sake of convenience we introduce the abbreviation
\begin{equation}\label{PN24}
 K_{\mu\nu}\equiv J_{\mu\nu}^{(0)}+\delta_{\mu\nu}\,\kappa_\mu^{(0)}
 ={\mathbbm J}_{\mu\nu}({\boldsymbol\lambda}^{(0)},\gamma_0 )-\delta_{\mu\nu}\,x_0
 \;,
\end{equation}
for all $\mu,\nu=1,\ldots,N$. The matrix (\ref{PN24}) defines a positively semi-definite operator $K$
with a two-dimensional kernel $\mbox{ker}(K)={\mathcal E}_0$ spanned by ${\mathbf 1}$ and ${\boldsymbol\xi}$.
Its $(N-2)$-dimensional range will be denoted by $\mbox{ran}(K)=\mbox{ker}(K)^\perp$.

%%%%%%%%%%%%%%%%%%%%%%%%%%%%%%%%%%%%%%%%%%%%%%%%%%%%%%%%%%%%%%%%%%%%%%%%%%%%%%%%%%%%%%%%%%%%%%%%%%%%%%%%%%%%%%%%%%%%
\subsubsection{Terms $O(t^2)$:}\label{sec:PS2}
%%%%%%%%%%%%%%%%%%%%%%%%%%%%%%%%%%%%%%%%%%%%%%%%%%%%%%%%%%%%%%%%%%%%%%%%%%%%%%%%%%%%%%%%%%%%%%%%%%%%%%%%%%%%%%%%%%%%

We obtain the second order terms of (\ref{PN6}):
\begin{equation}\label{PN25}
 \sum_\nu \left(
 J_{\mu\nu}^{(0)}\,{\mathbf s}_\nu^{(2)}+\Xi_{\mu\nu}\, {\mathbf s}_\nu^{(0)}
 \right)=
 -\kappa_\mu^{(0)}\,{\mathbf s}_\mu^{(2)}-\kappa_\mu^{(2)}\,{\mathbf s}_\mu^{(0)}
 \;,
\end{equation}
or, by means of (\ref{PN4c}),
\begin{equation}\label{PN26}
 \sum_\nu \left(
 J_{\mu\nu}^{(0)}\,x_\nu^{(2)}+\Xi_{\mu\nu}
 \right)=
 -\kappa_\mu^{(0)}\,x_\mu^{(2)}-\kappa_\mu^{(2)}
 \;,
\end{equation}
for $\mu=1,\ldots,N$.
Since the $x_\mu^{(2)}$ are already determined by (\ref{PN5a}),
we may view these equations as giving explicit expressions for the $\kappa_\mu^{(2)}$ for $\mu=1,\ldots,N$:
\begin{eqnarray}
\label{PN27a}
  \kappa_\mu^{(2)} &=& -\sum_\nu\left(J_{\mu\nu}^{(0)}+\delta_{\mu\nu}\,\kappa_\mu^{(0)}\right)y_\nu^{(2)}\;-(N-1) \\
  \label{PN27b}
  &\stackrel{(\ref{PN5a},\ref{PN24})}{=}&\frac{1}{2}\sum_\nu K_{\mu\nu}\,y_\nu^{(1)2}\;+(1-N)
  \;.
\end{eqnarray}
It follows that the vector $\kappa^{(2)}$ lies in the subspace spanned by $\mbox{ran(K)}$ and ${\mathbf 1}$ and hence is
orthogonal to ${\boldsymbol\xi}$ or, equivalently, to $y^{(1)}$:
\begin{equation}\label{PN28}
 \sum_\mu \kappa_\mu^{(2)}\,y_\mu^{(1)}=0
 \;.
\end{equation}
From (\ref{PN27b}) we may calculate the second order correction to the eigenvalue $x_0$ according to (\ref{PN4e}):
\begin{equation}\label{PN28a}
 x_2=-\frac{1}{N}\sum_\mu \kappa_\mu^{(2)}=-\frac{1}{2N}\left( \sum_{\mu\nu}K_{\mu\nu} y_\nu^{(1)2}\right) +N-1=N-1
 \;,
\end{equation}
since ${\mathbf 1}\in \mbox{ker}(K)$.

The second order correction to the magnetization reads
\begin{equation}\label{PN29}
 M^{(2)}=\sum_\mu x_\mu^{(2)}\stackrel{(\ref{PN5a})}{=}-\frac{1}{2}\sum_\mu y_\mu^{(1)2}\stackrel{(\ref{PN23})}{=}-\frac{1}{2}X^2
 \;.
\end{equation}

The analogous correction to the energy is obtained as
\begin{eqnarray}
\label{PN30a}
  E^{(2)} &=& \frac{1}{2}\sum_{\mu\nu} {\mathbbm J}_{\mu\nu}({\mathbf 0},0)
  \left(2{\mathbf s}_\mu^{(0)}\cdot{\mathbf s}_\nu^{(2)} +{\mathbf s}_\mu^{(1)}\cdot{\mathbf s}_\nu^{1)}\right)\\
  \label{PN30b}
  &\stackrel{(\ref{PN4c},\ref{PN5a})}{=}&
  \frac{1}{2}\sum_{\mu\nu} {\mathbbm J}_{\mu\nu}({\mathbf 0},0)
  \left(-y_\nu^{(1)2}+y_\mu^{(1)}\,y_\nu^{(1)}\right)\\
  \label{PN30c}
  &=&
  \frac{1}{2}\sum_{\mu\nu} J^{(h)}_{\mu\nu} \left(-y_\nu^{(1)2}+y_\mu^{(1)}\,y_\nu^{(1)}\right)\\
  \label{PN30d}
  &=&
  -\frac{1}{2}\sum_{\mu\nu} J^{(h)}_{\mu\nu}\,y_\nu^{(1)2}
  +\frac{1}{2}\sum_{\mu\nu} J^{(h)}_{\mu\nu} \,y_\mu^{(1)}\,y_\nu^{(1)}\\
  \label{PN30e}
  &\stackrel{(\ref{S4},\ref{S5},\ref{S7})}{=}& \frac{1}{2}\left( -j+j_{min}^{(h)}\right)\sum_\mu y_\mu^{(1)2}\\
  \label{PN30f}
 &\stackrel{(\ref{PN23})}{=}& \frac{1}{2}\left( -j+j_{min}^{(h)} \right)\,X^2
  \;.
 \end{eqnarray}
In Eq.~(\ref{PN30c}) we have used that the bracket in (\ref{PN30b}) vanishes for $\mu=\nu$ and hence the total expression is independent
of the diagonal elements of ${\mathbbm J}_{\mu\nu}({\mathbf 0},0)$. Especially, we may choose the diagonal elements
corresponding to the homogeneous gauge.

%%%%%%%%%%%%%%%%%%%%%%%%%%%%%%%%%%%%%%%%%%%%%%%%%%%%%%%%%%%%%%%%%%%%%%%%%%%%%%%%%%%%%%%%%%%%%%%%%%%%%%%%%%%%%%%%%%%%
\subsubsection{Terms $O(t^3)$:}\label{sec:PS3}
%%%%%%%%%%%%%%%%%%%%%%%%%%%%%%%%%%%%%%%%%%%%%%%%%%%%%%%%%%%%%%%%%%%%%%%%%%%%%%%%%%%%%%%%%%%%%%%%%%%%%%%%%%%%%%%%%%%%

The third order terms of (\ref{PN6}) are:
\begin{equation}\label{PN31}
 \sum_\nu \left(
 J_{\mu\nu}^{(0)}\,{\mathbf s}_\nu^{(3)}+\Xi_{\mu\nu}\, {\mathbf s}_\nu^{(1)}
 \right)=
 -\kappa_\mu^{(0)}\,{\mathbf s}_\mu^{(3)}-\kappa_\mu^{(2)}\,{\mathbf s}_\mu^{(1)}
 \;,
\end{equation}
or, using (\ref{PN4c}),
\begin{equation}\label{PN32}
 \sum_\nu \left(
 J_{\mu\nu}^{(0)}\,y_\nu^{(3)}+\Xi_{\mu\nu}\,y_\nu^{(1)}
 \right)=
 -\kappa_\mu^{(0)}\,y_\mu^{(3)}-\kappa_\mu^{(1)}
 \;,
\end{equation}
for $\mu=1,\ldots,N$.
By means of (\ref{PN24}) this can be brought into the form of an
(in general) inhomogeneous linear system of equations for the unknown $y_\nu^{(3)}$:
\begin{equation}\label{PN33}
 \sum_\nu K_{\mu\nu}\,y_\nu^{(3)} = \left( 1-\kappa_\mu^{(2)}\right) y_\mu^{(1)}\equiv u_\mu
 \;.
\end{equation}
This system is only solvable if the r.~h.~s.~lies in the range of $K$, i.e., $u\in\mbox{ran}(K)=\mbox{ker}(K)^\perp$.
We thus obtain the solvability conditions $u\perp {\mathbf 1}$ and $u\perp y^{(1)}$.
The first condition  follows from (\ref{PN5e}) and (\ref{PN28}). The second condition reads
\begin{equation}\label{PN34}
 \sum_\mu \left( 1-\kappa_\mu^{(2)}\right)\,y_\mu^{(1)2}=0
 \;.
\end{equation}
Obviously its validity depends of the value of $X$ that has not yet been determined.
So we may kill two birds with one stone by using (\ref{PN34}) to determine $X$:
\begin{eqnarray}
\label{PN35a}
  0 &=&  \sum_\mu \left( 1-\kappa_\mu^{(2)}\right)\,y_\mu^{(1)2} \\
  \label{PN35b}
    &\stackrel{(\ref{PN27b})}{=}& \sum_\mu\left(
    N-\frac{1}{2}\sum_\nu K_{\mu\nu}\,y_\nu^{(1)2}  \right)\,y_\mu^{(1)2} \\
    \label{PN35c}
   &\stackrel{(\ref{PN23})}{=}& N\,X^2 -\frac{1}{2}X^4\,\sum_{\mu\nu}K_{\mu\nu}\xi_\nu^2\,\xi_\mu^2  \\
   \label{PN35d}
   &=& X^2\,\left( N-\frac{1}{2}\,X^2\,k^2\right)
   \;,
\end{eqnarray}
with
\begin{equation}\label{PN36}
  k^2\equiv \sum_{\mu\nu}K_{\mu\nu}\,\xi_\nu^2\,\xi_\mu^2
  \;.
\end{equation}
$k^2\ge 0$ since it is defined as the expectation value of a positively semi-definite operator
and hence $k\equiv \sqrt{k^2}\ge 0$ is well-defined.
Then the second solvability condition equivalent to (\ref{PN35d}) yields
\begin{equation}\label{PN37}
 X=\frac{\sqrt{2N}}{k}
 \;.
\end{equation}
For the last equation it is required that $k^2> 0$. This can be proven as follows:
$k^2= \sum_{\mu\nu}K_{\mu\nu}\,\xi_\nu^2\,\xi_\mu^2=0$ is only possible if the vector with components
$(\xi_\mu^2)_{\mu=1,\ldots,N}$ lies in the linear span of ${\mathbf 1}$ and ${\boldsymbol\xi}$, that is
\begin{equation}\label{PN38}
  \xi_\mu^2 = \alpha\,\xi_\mu+\beta
\end{equation}
for two real numbers $\alpha$ and $\beta$ and all $\mu=1,\ldots,N$. Due to $\sum_\mu \xi_\mu^2=1$ and $\sum_\mu\xi_\mu=0$ we have $\beta=1$.
Further, $\alpha\neq 0$ since $\alpha=0$ would imply that all $\xi_\mu^2=1$ in contradiction to  $\sum_\mu \xi_\mu^2=1$
and the general condition $N\ge 3$.
Then the quadratic equation (\ref{PN38}) has the solutions
\begin{equation}\label{PN39}
 \xi_\mu = \frac{\alpha}{2}+\delta_\mu\,\sqrt{1+\frac{\alpha^2}{4}}
 \;,
\end{equation}
where $\delta_\mu=\pm 1$.
Let $\alpha>0$.
According to $\sum_\mu \xi_\mu=0$ not all $\delta_\mu$ can have the same sign.
Hence there exists at least one $\mu=1,\ldots,N$ with $\delta_\mu=+1$ such that
\begin{equation}\label{PN40 }
 \left| \xi_\mu\right|=\left|  \frac{\alpha}{2}+\sqrt{1+\frac{\alpha^2}{4}}\right|>1,
\end{equation}
in contradiction to $\sum_\mu \xi_\mu^2=1$.
If $\alpha<0$ we can argue analogously by choosing a $\delta_\mu=-1$.   \hfill$\Box$\\

For later purpose we consider
\begin{eqnarray}
\label{PN41a}
  \sum_\nu{\mathbbm J}_{\mu\nu}({\boldsymbol\lambda}^{(0)},\gamma_0)\,y_\nu^{(1)2}
  &\stackrel{(\ref{PN24})}{=}&
  \sum_\nu\left( K_{\mu\nu}+\delta_{\mu\nu}\,x_0\right)\,y_\nu^{(1)2} \\
  \label{PN41b}
  &\stackrel{(\ref{PN27b})}{=}& 2\left(N-1+\kappa_\mu^{(2)} \right) +x_0\,y_\mu^{(1)2}
  \;,
\end{eqnarray}
and further
\begin{eqnarray}
\label{PN42a}
  \sum_{\mu\nu}{\mathbbm J}_{\mu\nu}({\boldsymbol\lambda}^{(0)},\gamma_0)\,y_\nu^{(1)2}\,y_\mu^{(1)2}
  &\stackrel{(\ref{PN41b})}{=}&
  2\sum_{\mu}\left(N-1+\kappa_\mu^{(2)} \right)\,y_\mu^{(1)2} +x_0\sum_{\mu}y_\mu^{(1)4}\\
  \label{PN42b}
   &\stackrel{(\ref{PN34},\ref{PN23})}{=}& 2\,N\,X^2+x_0\sum_{\mu}y_\mu^{(1)4}
   \;.
\end{eqnarray}

%%%%%%%%%%%%%%%%%%%%%%%%%%%%%%%%%%%%%%%%%%%%%%%%%%%%%%%%%%%%%%%%%%%%%%%%%%%%%%%%%%%%%%%%%%%%%%%%%%%%%%%%%%%%%%%%%%%%
\subsubsection{Terms $O(t^4)$:}\label{sec:PS4}
%%%%%%%%%%%%%%%%%%%%%%%%%%%%%%%%%%%%%%%%%%%%%%%%%%%%%%%%%%%%%%%%%%%%%%%%%%%%%%%%%%%%%%%%%%%%%%%%%%%%%%%%%%%%%%%%%%%%

The fourth order terms of (\ref{PN6}) are:
\begin{equation}\label{PN43}
 \sum_\nu \left(
 J_{\mu\nu}^{(0)}\,{\mathbf s}_\nu^{(4)}+\Xi_{\mu\nu}\, {\mathbf s}_\nu^{(2)}
 \right)=
 -\kappa_\mu^{(0)}\,{\mathbf s}_\mu^{(4)}-\kappa_\mu^{(2)}\,{\mathbf s}_\mu^{(2)}-\kappa_\mu^{(4)}\,{\mathbf s}_\mu^{(0)}
 \;,
\end{equation}
or, using (\ref{PN4c}),
\begin{equation}\label{PN44}
 \sum_\nu \left(
 J_{\mu\nu}^{(0)}\,x_\nu^{(4)}+\Xi_{\mu\nu}\,x_\nu^{(2)}
 \right)=
 -\kappa_\mu^{(0)}\,x_\mu^{(4)}-\kappa_\mu^{(2)}\,x_\mu^{(2)}-\kappa_\mu^{(4)}
 \;,
\end{equation}
for $\mu=1,\ldots,N$. These equations can be used to calculate $\kappa_\mu^{(4)}$ for all $\mu=1,\ldots,N$:
\begin{eqnarray}
\label{PN45a}
 \kappa_\mu^{(4)} &\stackrel{(\ref{PN24},\ref{PN44})}{=}& -\sum_\nu K_{\mu\nu}\,x_\nu^{(4)}
-\sum_\nu \left( \Xi_{\mu\nu}+\kappa_\mu^{(2)}\,\delta_{\mu\nu} \right)\,x_\nu^{(2)}\\
\label{PN45b}
  &\stackrel{(\ref{PN5a},\ref{PN5b})}{=}& \sum_\nu K_{\mu\nu}\left(y_\nu^{(1)}\,y_\nu^{(3)}+\frac{1}{8}y_\nu^{(1)4} \right)
 +\frac{1}{2}\sum_\nu \left( \Xi_{\mu\nu}+\kappa_\mu^{(2)}\,\delta_{\mu\nu} \right)\,y_\nu^{(1)2}\\
 \label{PN45c}
 &\stackrel{(\ref{S2})}{=}&\sum_\nu K_{\mu\nu}\left(y_\nu^{(1)}\,y_\nu^{(3)}+\frac{1}{8}y_\nu^{(1)4} \right)
 +\frac{1}{2}\sum_\nu\left((1-\delta_{\mu\nu})y_\nu^{(1)2} \right)+\frac{1}{2}\,\kappa_\mu^{(2)}\,y_\mu^{(1)2}\\
 \label{PN45d}
  &\stackrel{(\ref{PN23})}{=}&\sum_\nu K_{\mu\nu}\left(y_\nu^{(1)}\,y_\nu^{(3)}+\frac{1}{8}y_\nu^{(1)4} \right)
  +\frac{1}{2}\left(X^2-y_\mu^{(1)2} \right)+\frac{1}{2}\,\kappa_\mu^{(2)}\,y_\mu^{(1)2}\\
  \label{PN45e}
   &=&\sum_\nu K_{\mu\nu}\left(y_\nu^{(1)}\,y_\nu^{(3)}+\frac{1}{8}y_\nu^{(1)4} \right)
  +\frac{1}{2}\left(X^2+\left(\kappa_\mu^{(2)}-1\right)y_\mu^{(1)2}\right)
  \;.
 \end{eqnarray}

From (\ref{PN45e}) we may calculate the fourth order correction to the eigenvalue $x_0$ according to (\ref{PN4e}):
\begin{eqnarray}\label{PN46a}
 x_4&=&-\frac{1}{N}\sum_\mu \kappa_\mu^{(4)}=-\frac{1}{2}\,X^2
 \;,
\end{eqnarray}
using (\ref{PN34}) and  ${\mathbf 1}\in \mbox{ker}(K)$.\\

The fourth order correction to the magnetization reads:
\begin{equation}\label{PN46}
 M^{(4)}=\sum_\mu x_\mu^{(4)}\stackrel{(\ref{PN5b})}{=}-\sum_\mu\left(y_\mu^{(1)}\,y_\mu^{(3)}+\frac{1}{8}y_\mu^{(1)4} \right)
 \;.
\end{equation}

For the analogous correction to the energy we obtain:
\begin{eqnarray}
\label{PN47a}
 E^{(4)} &=& \frac{1}{2}\sum_{\mu\nu}{\mathbbm J}_{\mu\nu}({\mathbf 0},0)\left(
 2\,{\mathbf s}_\mu^{(0)}\cdot{\mathbf s}_\nu^{(4)}+2\,{\mathbf s}_\mu^{(1)}\cdot{\mathbf s}_\nu^{(3)}+{\mathbf s}_\mu^{(2)}\cdot{\mathbf s}_\nu^{(2)}
 \right)\\
 \label{PN47b}
  &\stackrel{(\ref{PN5a},\ref{PN5b})}{=}&
  \frac{1}{2}\sum_{\mu\nu}{\mathbbm J}_{\mu\nu}({\mathbf 0},0)\left(
 -2\,y_\nu^{(1)}\,y_\nu^{(3)}-\frac{1}{4}\,y_\nu^{(1)4}+2\,y_\mu^{(1)}\,y_\nu^{(3)}+\frac{1}{4}\,y_\mu^{(1)2}\,y_\nu^{(1)2}
 \right)
 \;.
 \end{eqnarray}
Since the bracket in the last equation vanishes for $\mu=\nu$ we may add arbitrary diagonal elements to ${\mathbbm J}_{\mu\nu}({\mathbf 0},0)$
without changing the total value of $E^{(4)}$. In particular, we may choose the homogeneous gauge of the J-matrix thus obtaining:
\begin{eqnarray}
\label{PN48a}
 E^{(4)} &=&
  \frac{1}{2}\sum_{\mu\nu}J^{(h)}_{\mu\nu}\left(
 -2\,y_\nu^{(1)}\,y_\nu^{(3)}-\frac{1}{4}\,y_\nu^{(1)4}+2\,y_\mu^{(1)}\,y_\nu^{(3)}+\frac{1}{4}\,y_\mu^{(1)2}\,y_\nu^{(1)2}
 \right)\\
 \label{PN48b}
 &\stackrel{(\ref{S4},\ref{S5},\ref{S1})}{=}& -j \sum_\nu \left( y_\nu^{(1)}\,y_\nu^{(3)}+\frac{1}{8}\,y_\nu^{(1)4}\right)+
 \sum_{\mu\nu}\left({\mathbbm J}_{\mu\nu}({\boldsymbol\lambda}^{(0)},\gamma_0)-\gamma_0\,\Xi_{\mu\nu} \right)
 \left( y_\mu^{(1)}\,y_\nu^{(3)}+\frac{1}{8}\,y_\mu^{(1)2}\,y_\nu^{(1)2}\right)\\
  \nonumber
 &\stackrel{(\ref{PN46},\ref{PN22},\ref{S4},\ref{S2})}{=}&
 j\,M^{(4)}+x_0\,\sum_\nu y_\nu^{(1)}\,y_\nu^{(3)}+\gamma_0\,\sum_\nu y_\nu^{(1)}\,y_\nu^{(3)}+
  \frac{1}{8}\sum_{\mu\nu}{\mathbbm J}_{\mu\nu}({\boldsymbol\lambda}^{(0)},\gamma_0)\,y_\mu^{(1)2}\,y_\nu^{(1)2}\\
  \label{PN48c}
 &&
 -\frac{\gamma_0}{8}
 \left[
 \left(\sum_\mu y_\mu^{(1)2} \right)\,\left(\sum_\nu y_\nu^{(1)2} \right)-\sum_\mu y_\mu^{(1)4}
 \right]\\
 \nonumber
 &\stackrel{(\ref{PN23})}{=}&
 j M^{(4)}+x_0\sum_\nu y_\nu^{(1)}y_\nu^{(3)}+\gamma_0\sum_\nu \left( y_\nu^{(1)}y_\nu^{(3)}+\frac{1}{8}y_\nu^{(1)4}\right)+
 \frac{1}{8}\sum_{\mu\nu}{\mathbbm J}_{\mu\nu}({\boldsymbol\lambda}^{(0)},\gamma_0)y_\mu^{(1)2}y_\nu^{(1)2} -\frac{\gamma_0}{8}X^4\\
 \label{PN48d}
 &&\\
&\stackrel{(\ref{PN42b},\ref{S12a})}{=}&j\, M^{(4)}+(x_0+\gamma_0)\,\sum_\nu \left(y_\nu^{(1)}y_\nu^{(3)}+\frac{1}{8}y_\nu^{(1)4}\right)
+\frac{N}{4}\,X^2 +\frac{j-j_{min}^{(h)}}{8N}\,X^4\\
\label{PN48e}
&\stackrel{(\ref{PN46},\ref{S12a})}{=}&\left(j-j_{min}^{(h)}\right)\,M^{(4)}+\frac{j-j_{min}^{(h)}}{8N}\,X^4+\frac{N}{4}\,X^2
\;.
\end{eqnarray}

%%%%%%%%%%%%%%%%%%%%%%%%%%%%%%%%%%%%%%%%%%%%%%%%%%%%%%%%%%%%%%%%%%%%%%%%%%%%%%%%%%%%%%%%%%%%%%%%%%%%%%%%%%%%%%%%%%%%%%%%%%%%%%%%%%%%%%%%%%
\subsubsection{Saturation susceptibility}\label{sec:SS}
%%%%%%%%%%%%%%%%%%%%%%%%%%%%%%%%%%%%%%%%%%%%%%%%%%%%%%%%%%%%%%%%%%%%%%%%%%%%%%%%%%%%%%%%%%%%%%%%%%%%%%%%%%%%%%%%%%%%%%%%%%%%%%%%%%%%%%%%%%

We will use the series coefficients of $M(t)$ and $E(t)$ calculated in the preceding subsections to determine the leading coefficient of the
susceptibility. To this end we first consider the series expansion of the magnetic field
\begin{eqnarray}
\label{SS1a}
 B(t) &=& \frac{\partial E/\partial t}{\partial M/\partial t} \\
 \label{SS1b}
   &=& \frac{2\, E^{(2)}\, t+ 4\, E^{(4)}\,t^3+\ldots}{2\, M^{(2)}\, t+ 4\, M^{(4)\,}t^3+\ldots}\\
   \label{SS1c}
   &=& \frac{E^{(2)}}{M^{(2)}}+\frac{2\left(E^{(4)}M^{(2)}-E^{(2)}M^{(4)} \right)}{M^{(2)2}}\,t^2+\ldots
 \end{eqnarray}
This yields the saturation field
\begin{equation}\label{SS2}
 B_{sat}= \lim_{t\rightarrow 0}B(t)=\frac{E^{(2)}}{M^{(2)}}\stackrel{(\ref{PN30f},\ref{PN29})}{=}
  \frac{\frac{1}{2}\left( -j+j_{min}^{(h)} \right)\,X^2}{-\frac{1}{2}X^2}=j-j_{min}^{(h)}
  \;.
\end{equation}
This result is in accordance with (\ref{HZS1}).\\

Next we consider the series representation of the susceptibility
\begin{eqnarray}
\label{SS3a}
  \chi(t)&=& \frac{\partial M/\partial t}{\partial B/\partial t} \\
  &\stackrel{(\ref{SS1c})}{=}&\frac{2\, M^{(2)}\, t+ 4\, M^{(4)\,}t^3+\ldots}{\frac{4t}{M^{(2)2}}\left(E^{(4)}M^{(2)}-E^{(2)}M^{(4)} \right)+\ldots}
  \;.
\end{eqnarray}
This yields the saturation susceptibility
\begin{eqnarray}\label{SS4a}
 \chi_0&\equiv& \lim_{t\rightarrow 0}\chi(t) =\frac{M^{(2)3}}{2\left(E^{(4)}M^{(2)}-E^{(2)}M^{(4)} \right)}\\
 \label{SS4b}
 &=&\frac{\left(-\frac{1}{2}X^2\right)^3}{2\left(B_{sat}M^{(4)}+\frac{B_{sat}X^4}{8N}+\frac{N}{4}X^4 \right)
 \left(-\frac{1}{2}X^2\right)-2\left(-\frac{1}{2} B_{sat}X^2\right)\left(M^{(4)}\right)}\\
 \label{SS4c}
 &=& \frac{N\,X^2}{B_{sat}\,X^2+2\,N^2}\\
 \label{SS4d}
 &\stackrel{(\ref{PN37})}{=}& \frac{N}{B_{sat}+N k^2}
 \;,
\end{eqnarray}
which represents a central result of the present paper. In Eq.~(\ref{SS4b}) we have inserted the
previous results for $M^{(2)},\,E^{(2)},$ and $E^{(4)}$, see (\ref{PN29}), (\ref{PN30c}), and (\ref{PN48e}),
together with (\ref{SS2}).

%%%%%%%%%%%%%%%%%%%%%%%%%%%%%%%%%%%%%%%%%%%%%%%%%%%%%%%%%%%%%%%%%%%%%%%%%%%%%%%%%%%%%%%%%%%%%%%%%%%%%%%%%%%%%%%%%%%%%%%%%%%%%%%%%%%%%%%%%%
\subsection{Three-dimensional ground states} \label{sec:TGS}
%%%%%%%%%%%%%%%%%%%%%%%%%%%%%%%%%%%%%%%%%%%%%%%%%%%%%%%%%%%%%%%%%%%%%%%%%%%%%%%%%%%%%%%%%%%%%%%%%%%%%%%%%%%%%%%%%%%%%%%%%%%%%%%%%%%%%%%%%%
We assume that the eigenspace ${\mathcal E}_0$ of ${\mathbbm J}_{\mu\nu}({\boldsymbol{\lambda}}^{(0)},\gamma_0)$
corresponding to the lowest eigenvalue $x_0$
is three-dimensional and hence the subspace ${\mathcal E}_1$ according to (\ref{NP1}) will be two-dimensional.
Let $({\boldsymbol\xi}^{(1)},{\boldsymbol\xi}^{(2)})$  be a fixed orthonormal basis in ${\mathcal E}_1$.

%%%%%%%%%%%%%%%%%%%%%%%%%%%%%%%%%%%%%%%%%%%%%%%%%%%%%%%%%%%%%%%%%%%%%%%%%%%%%%%%%%%%%%%%%%%%%%%%%%%%%%%%%%%%%%%%%%%%%%%%%%%%%%%%%%%%%%%%%%
\subsubsection{Notations and first results}\label{sec:TGSPN}
%%%%%%%%%%%%%%%%%%%%%%%%%%%%%%%%%%%%%%%%%%%%%%%%%%%%%%%%%%%%%%%%%%%%%%%%%%%%%%%%%%%%%%%%%%%%%%%%%%%%%%%%%%%%%%%%%%%%%%%%%%%%%%%%%%%%%%%%%%

Similarly as in Section  \ref{sec:PN} we consider the one-parameter families
\begin{eqnarray}\label{TGS4a}
 J_{\mu\nu}(t)&=&J_{\mu\nu}^{(0)}+t^2\,\Xi_{\mu\nu}\;,\\
\label{TGS4b}
{\mathbf s}_\mu(t)&=& \sum_{n=0,1,2,\ldots}t^n\,{\mathbf s}_\mu^{(n)}\;,\\
\label{TGS4c}
&=&{1\choose 0}+t\,{0\choose {\mathbf y}_\mu^{(1)}}+t^2\,{ x_\mu^{(2)}\choose 0}+t^3\,{0\choose {\mathbf y}_\mu^{(3)}}+t^4\,{ x_\mu^{(4)}\choose 0}+\ldots\;,\\
\label{TGS4d}
\kappa_\mu(t)&=& \sum_{n=0,2,4,\ldots}t^n\,{\kappa}_\mu^{(n)}\;,\\
\label{TGS4e}
x(t)&=&-\frac{1}{N}\sum_\mu \kappa_\mu(t)= \sum_{n=0,2,4,\ldots}t^n\,x_n
\;,
\end{eqnarray}
for $\mu=1,\ldots,N$. However, in this section the vectors ${\mathbf y}_\mu^{(n)}$ for odd $n$ are assumed to be two-dimensional,
${\mathbf y}_\mu^{(n)}\in{\mathbbm R}^2$,
and their components are designated $y_\mu^{(n,a)}$ for $a=1,2$.
For fixed $n$ we may view the $y_\mu^{(n,a)}$ as the entries of an $N\times 2$-matrix ${\mathbf y}^{(n)}$ with $N$ rows ${\mathbf y}_\mu^{(n)}$
and two columns ${\mathbf y}^{(n,a)}$.

The condition $\|{\mathbf s}_\mu(t)\|=1$ for all $\mu=1,\ldots,N$
entails an infinite number of identities for the $x_\mu^{(n)}$ , the first two of which read
\begin{eqnarray}
\label{TGS5a}
 x_\mu^{(2)} &=& -\frac{1}{2}\,{\mathbf y}_\mu^{(1)}\cdot{\mathbf y}_\mu^{(1)},\\
\label{TGS5b}
 x_\mu^{(4)} &=& -\left( {\mathbf y}_\mu^{(1)}\cdot{\mathbf y}_\mu^{(3)}+\frac{1}{8}\left({\mathbf y}_\mu^{(1)}\cdot{\mathbf y}_\mu^{(1)} \right)^2\right)
\;.
\end{eqnarray}
In the ground state configuration the total spin ${\mathbf S}(t)$ will point into the direction ${1\choose {\mathbf 0}}$ of the field and hence
\begin{equation}\label{TGS5c}
 {\mathbf S}(t) = \sum_\mu {\mathbf s}_\mu(t)\equiv {M(t)\choose {\mathbf 0}}
 \;,
\end{equation}
such that
\begin{equation}\label{TGS5d}
M(t)=N+t^2 M^{(2)}+t^4\,M^{(4)}+\ldots
\stackrel{(\ref{TGS4c},\ref{TGS5c})}{=}
\sum_{n=0,2,4,\ldots} t^n\;\; \sum_\mu x_\mu^{(n)}
\;.
\end{equation}
We note that Eqs.~(\ref{TGS5c}) and (\ref{TGS4c}) imply
\begin{equation}\label{TGS5e}
 \sum_\mu {\mathbf y}_\mu^{(n)} = {\mathbf 0} \quad \mbox{ for all odd } n
 \;.
\end{equation}
Further we consider the energy (without the auxiliary uniform coupling)
\begin{equation}\label{TGS5f}
 E(t)=\frac{1}{2}\sum_{\mu\nu} {\mathbbm J}_{\mu\nu}({\mathbf 0},0)\,{\mathbf s}_\mu(t)\cdot {\mathbf s}_\mu(t)
 = E^{(0)}+t^2 E^{(2)}+t^4 E^{(4)}+\ldots
\end{equation}
The $t$-series for $M(t)$ and $E(t)$ contain only even terms since the scalar product of two terms of different parity in (\ref{TGS4b}) vanishes.

%%%%%%%%%%%%%%%%%%%%%%%%%%%%%%%%%%%%%%%%%%%%%%%%%%%%%%%%%%%%%%%%%%%%%%%%%%%%%%%%%%%%%%%%%%%%%%%%%%%%%%%%%%%%%%%%%%%%%%%%%%%%%%%%%%%%%%%%%%
\subsubsection{Perturbation series}\label{sec:TGSPS}
%%%%%%%%%%%%%%%%%%%%%%%%%%%%%%%%%%%%%%%%%%%%%%%%%%%%%%%%%%%%%%%%%%%%%%%%%%%%%%%%%%%%%%%%%%%%%%%%%%%%%%%%%%%%%%%%%%%%%%%%%%%%%%%%%%%%%%%%%%

The procedure is analogous to that of section \ref{sec:CGS} and hence we will only mention
those equations and results that are essentially different.
Especially the terms of $0^{th}$ order in $t$ are identical to those of section \ref{sec:S}.

%%%%%%%%%%%%%%%%%%%%%%%%%%%%%%%%%%%%%%%%%%%%%%%%%%%%%%%%%%%%%%%%%%%%%%%%%%%%%%%%%%%%%%%%%%%%%%%%%%%%%%%%%%%%%%%%%%%%
\subsubsection{Terms $O(t^1)$:}\label{sec:TGST1}
%%%%%%%%%%%%%%%%%%%%%%%%%%%%%%%%%%%%%%%%%%%%%%%%%%%%%%%%%%%%%%%%%%%%%%%%%%%%%%%%%%%%%%%%%%%%%%%%%%%%%%%%%%%%%%%%%%%%

The $t$-linear terms of (\ref{PN6}) read:
\begin{equation}\label{PN20}
  \sum_\nu J_{\mu\nu}^{(0)}\,{\mathbf s}_\nu^{(1)}=-\kappa_\mu^{(0)}\,{\mathbf s}_\mu^{(1)}
  \;.
\end{equation}
Using (\ref{TGS4c}) this means that
\begin{equation}\label{TGS21}
  \sum_\nu J_{\mu\nu}^{(0)}\,{\mathbf y}_\nu^{(1)}=-\kappa_\mu^{(0)}\,{\mathbf y}_\mu^{(1)}
  \;,
\end{equation}
or, due to (\ref{T4}) and (\ref{S12a}),
\begin{equation}\label{TGS22}
  \sum_\nu {\mathbbm J}_{\mu\nu}({\boldsymbol\lambda}^{(0)},\gamma_0)\,{\mathbf y}_\nu^{(1)}=x_0\,{\mathbf y}_\mu^{(1)}
  \;.
\end{equation}
Hence the columns ${\mathbf y}^{(1,a)},\;a=1,2$, of the matrix ${\mathbf y}^{(1)}$ are eigenvectors of
${\mathbbm J}({\boldsymbol\lambda}^{(0)},\gamma_0)$ corresponding to its lowest eigenvalue $x_0$.
According to (\ref{TGS5e}) these eigenvectors  ${\mathbf y}^{(1,a)}$ are orthogonal to ${\mathbf 1}$
and hence lie in ${\mathcal E}_1$. They can hence be expanded into the orthonormal basis $({\boldsymbol\xi}^{(1)},{\boldsymbol\xi}^{(2)})$:
\begin{eqnarray}\label{TGS22a}
  {\mathbf y}_\mu^{(1,1)}&=&\alpha_{11}\,{\boldsymbol\xi}_\mu^{(1)}+\alpha_{12}\,{\boldsymbol\xi}_\mu^{(2)},\\
  \label{TGS22b}
  {\mathbf y}_\mu^{(1,2)}&=&\alpha_{21}\,{\boldsymbol\xi}_\mu^{(1)}+\alpha_{22}\,{\boldsymbol\xi}_\mu^{(2)}
  \;,
\end{eqnarray}
for all $\mu=1,\ldots,N$. The $\alpha_{ij}$ can be viewed as the coefficients of a $2\times 2$-matrix
${\boldsymbol\alpha}$. It is only unique up to an arbitrary rotation/reflection in the two-dimensional space
${\mathcal E}_1$. This freedom can be used to additionally require that ${\boldsymbol\alpha}$ is symmetric
and positively definite, ${\boldsymbol\alpha}>0$.
In fact,
let ${\boldsymbol\alpha}= \left({\boldsymbol\alpha}\,{\boldsymbol\alpha}^\top \right)^{1/2}\,R$
be the polar decomposition of ${\boldsymbol\alpha}$ with $R\in O(2)$,
then
${\boldsymbol\alpha}\,R^{-1}= \left({\boldsymbol\alpha}\,{\boldsymbol\alpha}^\top \right)^{1/2}$
will be positively semi-definite. The stronger requirement ${\boldsymbol\alpha}>0$ follows from the condition
of a proper two-dimensional vector ${\mathbf y}^{(1)}$.\\

 We will determine ${\boldsymbol\alpha}$ below.

%%%%%%%%%%%%%%%%%%%%%%%%%%%%%%%%%%%%%%%%%%%%%%%%%%%%%%%%%%%%%%%%%%%%%%%%%%%%%%%%%%%%%%%%%%%%%%%%%%%%%%%%%%%%%%%%%%%%
\subsubsection{Terms $O(t^2)$:}\label{sec:TGST2}
%%%%%%%%%%%%%%%%%%%%%%%%%%%%%%%%%%%%%%%%%%%%%%%%%%%%%%%%%%%%%%%%%%%%%%%%%%%%%%%%%%%%%%%%%%%%%%%%%%%%%%%%%%%%%%%%%%%%

We obtain the second order terms of (\ref{PN6}):
\begin{equation}\label{TGS25}
 \sum_\nu \left(
 J_{\mu\nu}^{(0)}\,{\mathbf s}_\nu^{(2)}+\Xi_{\mu\nu}\, {\mathbf s}_\nu^{(0)}
 \right)=
 -\kappa_\mu^{(0)}\,{\mathbf s}_\mu^{(2)}-\kappa_\mu^{(2)}\,{\mathbf s}_\mu^{(0)}
 \;,
\end{equation}
or, by means of (\ref{TGS4c}),
\begin{equation}\label{TGS26}
 \sum_\nu \left(
 J_{\mu\nu}^{(0)}\,x_\nu^{(2)}+\Xi_{\mu\nu}
 \right)=
 -\kappa_\mu^{(0)}\,x_\mu^{(2)}-\kappa_\mu^{(2)}
 \;,
\end{equation}
for $\mu=1,\ldots,N$.
Since the $x_\mu^{(2)}$ are already determined by (\ref{TGS5a}),
we may view these equations as explicit expressions for the $\kappa_\mu^{(2)}$ for $\mu=1,\ldots,N$:
\begin{eqnarray}
\label{TGS27a}
  \kappa_\mu^{(2)} &=& -\sum_\nu\left(J_{\mu\nu}^{(0)}+\delta_{\mu\nu}\,\kappa_\mu^{(0)}\right)x_\nu^{(2)}\;-(N-1) \\
  \label{TGS27b}
  &\stackrel{(\ref{TGS5a},\ref{PN24})}{=}&\frac{1}{2}\sum_\nu K_{\mu\nu}\,{\mathbf y}_\mu^{(1)}\cdot{\mathbf y}_\mu^{(1)}\;+(1-N)
  \;.
\end{eqnarray}
It follows that the vector $\kappa^{(2)}$ lies in the subspace spanned by $\mbox{ran(K)}$ and ${\mathbf 1}$ and hence is
orthogonal to ${\mathcal E}_1$ or, equivalently, to ${\boldsymbol\xi}^{(1)}$ and ${\boldsymbol\xi}^{(2)}$:
\begin{equation}\label{TGS28}
 \sum_\mu \kappa_\mu^{(2)}\,\xi_\mu^{(1)}=\sum_\mu \kappa_\mu^{(2)}\,\xi_\mu^{(2)}=0
 \;.
\end{equation}

From (\ref{TGS27b}) we may calculate the second order correction to the eigenvalue $x_0$ according to (\ref{PN4e}):
\begin{equation}\label{TGS28a}
 x_2=-\frac{1}{N}\sum_\mu \kappa_\mu^{(2)}=-\frac{1}{2N}\left( \sum_{\mu\nu}K_{\mu\nu}\, {\mathbf y}_\mu^{(1)}\cdot{\mathbf y}_\mu^{(1)}\right) +N-1=N-1
 \;,
\end{equation}
since ${\mathbf 1}\in \mbox{ker}(K)$.

The second order correction to the magnetization reads
\begin{equation}\label{TGS29}
 M^{(2)}=\sum_\mu x_\mu^{(2)}\stackrel{(\ref{TGS5a})}{=}-\frac{1}{2}\sum_\mu {\mathbf y}_\mu^{(1)}\cdot{\mathbf y}_\mu^{(1)}
 \;.
\end{equation}

The analogous correction to the energy is obtained as
\begin{eqnarray}
\label{TGS30a}
  E^{(2)} &=& \frac{1}{2}\sum_{\mu\nu} {\mathbbm J}_{\mu\nu}({\mathbf 0},0)
  \left(2{\mathbf s}_\mu^{(0)}\cdot{\mathbf s}_\nu^{(2)} +{\mathbf s}_\mu^{(1)}\cdot{\mathbf s}_\nu^{1)}\right)\\
  \label{TGS30b}
  &\stackrel{(\ref{TGS4c},\ref{TGS5a})}{=}&
  \frac{1}{2}\sum_{\mu\nu} {\mathbbm J}_{\mu\nu}({\mathbf 0},0)
  \left(-{\mathbf y}_\mu^{(1)}\cdot{\mathbf y}_\mu^{(1)}+{\mathbf y}_\mu^{(1)}\cdot{\mathbf y}_\nu^{(1)}\right)\\
  \label{TGS30c}
  &=&
  \frac{1}{2}\sum_{\mu\nu} J^{(h)}_{\mu\nu}  \left(-{\mathbf y}_\mu^{(1)}\cdot{\mathbf y}_\mu^{(1)}+{\mathbf y}_\mu^{(1)}\cdot{\mathbf y}_\nu^{(1)}\right)\\
  \label{TGS30d}
  &=&
  -\frac{1}{2}\sum_{\mu\nu} J^{(h)}_{\mu\nu}\,{\mathbf y}_\mu^{(1)}\cdot{\mathbf y}_\mu^{(1)}
  +\frac{1}{2}\sum_{\mu\nu} J^{(h)}_{\mu\nu} \,{\mathbf y}_\mu^{(1)}\cdot{\mathbf y}_\nu^{(1)}\\
  \label{TGS30e}
  &\stackrel{(\ref{S4},\ref{S5},\ref{S7})}{=}& \frac{1}{2}\left( -j+j_{min}^{(h)}\right)\sum_\mu \,{\mathbf y}_\mu^{(1)}\cdot{\mathbf y}_\mu^{(1)}\\
  \label{TGS30f}
 &\stackrel{(\ref{TGS29})}{=}& \left( j-j_{min}^{(h)} \right)\,M^{(2)}
  \;.
 \end{eqnarray}
In Eq.~(\ref{TGS30c}) we have used that the bracket in (\ref{TGS30b}) vanishes for $\mu=\nu$ and hence the total expression is independent
of the matrix' diagonal.

%%%%%%%%%%%%%%%%%%%%%%%%%%%%%%%%%%%%%%%%%%%%%%%%%%%%%%%%%%%%%%%%%%%%%%%%%%%%%%%%%%%%%%%%%%%%%%%%%%%%%%%%%%%%%%%%%%%%
\subsubsection{Terms $O(t^3)$:}\label{sec:TGST3}
%%%%%%%%%%%%%%%%%%%%%%%%%%%%%%%%%%%%%%%%%%%%%%%%%%%%%%%%%%%%%%%%%%%%%%%%%%%%%%%%%%%%%%%%%%%%%%%%%%%%%%%%%%%%%%%%%%%%

The third order terms of (\ref{PN6}) are:
\begin{equation}\label{TGS31}
 \sum_\nu \left(
 J_{\mu\nu}^{(0)}\,{\mathbf s}_\nu^{(3)}+\Xi_{\mu\nu}\, {\mathbf s}_\nu^{(1)}
 \right)=
 -\kappa_\mu^{(0)}\,{\mathbf s}_\mu^{(3)}-\kappa_\mu^{(2)}\,{\mathbf s}_\mu^{(1)}
 \;,
\end{equation}
or, using (\ref{TGS4c}),
\begin{equation}\label{TGS32}
 \sum_\nu \left(
 J_{\mu\nu}^{(0)}\,{\mathbf y}_\nu^{(3)}+\Xi_{\mu\nu}\,{\mathbf y}_\nu^{(1)}
 \right)=
 -\kappa_\mu^{(0)}\,{\mathbf y}_\mu^{(3)}-\kappa_\mu^{(1)}
 \;,
\end{equation}
for $\mu=1,\ldots,N$.
By means of (\ref{PN24}) this can be brought into the form of an
(in general) inhomogeneous linear system of equations for the unknown ${\mathbf y}_\nu^{(3)}$:
\begin{equation}\label{TGS33}
 \sum_\nu K_{\mu\nu}\,{\mathbf y}_\nu^{(3)} = \left( 1-\kappa_\mu^{(2)}\right) {\mathbf y}_\mu^{(1)}\equiv {\mathbf u}_\mu
 \;.
\end{equation}
This system is only solvable if the r.~h.~s.~lies in the range of $K$, i.e., ${\mathbf u}^{(a)}\in\mbox{ran}(K)=\mbox{ker}(K)^\perp$
for $a=1,2$. Especially, the solvability condition implies
\begin{equation}\label{TGS33a}
 \sum_\mu \left( 1-\kappa_\mu^{(2)}\right) {\mathbf y}_\mu^{(1)}\cdot{\mathbf y}_\mu^{(1)}=0
 \;.
\end{equation}

More generally, we obtain the solvability conditions ${\mathbf u}^{(a)}\perp {\mathbf 1}$ and ${\mathbf u}^{(a)}\perp {\boldsymbol\xi}^{(b)}$
for all $a,b=1,2$.
The first conditions  follow from (\ref{TGS5e}) and (\ref{TGS28}). Using (\ref{TGS27b}), the second group of conditions is
equivalent to
\begin{equation}\label{TGS34}
0=\sum_\mu \left(2\,N\,{\mathbf y}_\mu^{(1,a)}-\sum_\nu K_{\mu\nu}\,{\mathbf y}_\nu^{(1)}\cdot {\mathbf y}_\nu^{(1)}\,{\mathbf y}_\mu^{(1,a)}\right){\boldsymbol\xi}_\mu^{(b)}
 \;,
\end{equation}
for all $a,b=1,2$. Upon expanding the ${\mathbf y}_\mu^{(1,a)}$ in terms of the ${\boldsymbol\xi}_\mu^{(b)}$ via (\ref{TGS22a}) and (\ref{TGS22b})
we thus obtain four polynomial equations of third order for the four unknown $\alpha_{ab}$. In order to write these
equations in concise form we introduce the three vectors ${\mathbf q}^{(i)},\;i=1,2,3,$ with components
\begin{equation}\label{TGS35}
  {\mathbf q}_\mu^{(1)}\equiv {\boldsymbol\xi}_\mu^{(1)2},
  \;{\mathbf q}_\mu^{(2)}\equiv {\boldsymbol\xi}_\mu^{(1)}\,{\boldsymbol\xi}_\mu^{(2)},
  \;{\mathbf q}_\mu^{(3)}\equiv {\boldsymbol\xi}_\mu^{(2)2},
\end{equation}
for $\mu=1,\ldots,N$ and define
\begin{equation}\label{TGS36}
  k_{ij}\equiv \sum_{\mu\nu}{\mathbf q}_\mu^{(i)}\,K_{\mu\nu}\,{\mathbf q}_\nu^{(j)}
  \;,
\end{equation}
for $1\le i,j\le 3$, where $k_{ij}=k_{ji}$ follows from the symmetry of $K$.
Then the four polynomial equations assume the form
\begin{eqnarray}
\nonumber
  2 N \alpha_{11} &=& \alpha _{11} \left(\alpha _{11}^2+\alpha _{21}^2\right)k_{11}
    +\alpha _{12} \left(\alpha _{11}^2+\alpha _{21}^2\right) k_{12}
    +\alpha _{11} \left(\alpha _{12}^2+\alpha _{22}^2\right) k_{13}
    + \alpha _{12} \left(\alpha _{12}^2+\alpha _{22}^2\right) k_{23} \\
    \label{TGS37a}
    &&+2 \alpha_{11} \left(\alpha _{11} \alpha _{12}+\alpha _{21} \alpha _{22}\right) k_{12}
    +2 \alpha _{12} \left(\alpha _{11} \alpha _{12}+\alpha _{21} \alpha _{22}\right)k_{22}\;,\\
    \nonumber
    2 N \alpha _{12} &=& \alpha _{11} \left(\alpha _{11}^2+\alpha _{21}^2\right)k_{12}
    +\alpha _{12} \left(\alpha _{11}^2+\alpha _{21}^2\right) k_{13}
    +\alpha _{11} \left(\alpha _{12}^2+\alpha _{22}^2\right) k_{23}
    +\alpha_{12} \left(\alpha _{12}^2+\alpha _{22}^2\right) k_{33}\\
     \label{TGS37b}
    &&  +2 \alpha_{11} \left(\alpha _{11} \alpha _{12}+\alpha _{21} \alpha _{22}\right) k_{22}
    +2 \alpha_{12} \left(\alpha _{11} \alpha _{12}+\alpha _{21} \alpha _{22}\right) k_{23}\;,\\
    \nonumber
    2N \alpha _{21}&=&\alpha _{21} \left(\alpha _{11}^2+\alpha _{21}^2\right) k_{11}
    +\alpha _{22} \left(\alpha _{11}^2+\alpha _{21}^2\right) k_{12}
    +\alpha _{21} \left(\alpha _{12}^2+\alpha _{22}^2\right) k_{13}
    +\alpha_{22} \left(\alpha _{12}^2+\alpha _{22}^2\right) k_{23}\\
     \label{TGS37c}
    && +2 \alpha_{21} \left(\alpha _{11} \alpha _{12}+\alpha _{21} \alpha _{22}\right) k_{12}
    +2 \alpha _{22} \left(\alpha _{11} \alpha _{12}+\alpha _{21} \alpha _{22}\right)k_{22}\;,\\
    \nonumber
    2 N \alpha _{22}&=&\alpha _{21} \left(\alpha _{11}^2+\alpha _{21}^2\right) k_{12}
    +\alpha _{22} \left(\alpha _{11}^2+\alpha _{21}^2\right) k_{13}
   +\alpha _{21} \left(\alpha _{12}^2+\alpha _{22}^2\right) k_{23}
   +\alpha_{22} \left(\alpha _{12}^2+\alpha _{22}^2\right) k_{33}\\
   \label{TGS37d}
   &&+2 \alpha_{21} \left(\alpha _{11} \alpha _{12}+\alpha _{21} \alpha _{22}\right) k_{22}
    +2 \alpha _{22} \left(\alpha _{11} \alpha _{12}+\alpha _{21} \alpha _{22}\right)k_{23}
    \;.
  \end{eqnarray}
These equations could be further simplified, but this appears superfluous as they surprisingly can be directly
solved using computer-algebraic means, if we add the equation $\alpha_{12}=\alpha_{21}$ considered above
and use concrete numbers for the $k_{ij}$ calculated for the particular spin system under consideration.
This will be demonstrated below for the example in Section \ref{sec:IH}. Note, that the above system
(\ref{TGS37a}) - (\ref{TGS37d}) is independent of $N$, only the calculation of the $k_{ij}$ may become
more cumbersome for large $N$. Typically, a computer-algebraic system would yield a finite number of
solutions that can, however, be boiled down to a single solution by using the condition
${\boldsymbol\alpha}>0$ considered above.

Hence we will proceed by assuming that a unique solution of (\ref{TGS37a}) - (\ref{TGS37d}) together with $\alpha_{12}=\alpha_{21}$
exists and leave it to the particular case how to concretely calculate ${\boldsymbol\alpha}$. We observe that the quadratic correction to the
magnetization  can be expressed in terms of the $\alpha_{ij}$:
\begin{equation}\label{TGS38}
 M^{(2)}\stackrel{(\ref{TGS29})}{=}-\frac{1}{2}\sum_\mu {\mathbf y}_\mu^{(1)}\cdot{\mathbf y}_\mu^{(1)}
 \stackrel{(\ref{TGS22a},\ref{TGS22b})}{=}-\frac{1}{2}\,\mbox{Tr } \left( {\boldsymbol\alpha}\,{\boldsymbol\alpha}^\top\right)
 = -\frac{1}{2}\left( \alpha_{11}^2+2 \alpha_{12}^2+\alpha_{22}^2\right)
 \;.
\end{equation}

For later purpose we consider
\begin{eqnarray}
\label{TGS41a}
  \sum_\nu{\mathbbm J}_{\mu\nu}({\boldsymbol\lambda}^{(0)},\gamma_0)\,{\mathbf y}_\nu^{(1)}\cdot{\mathbf y}_\nu^{(1)}
  &\stackrel{(\ref{PN24})}{=}&
  \sum_\nu\left( K_{\mu\nu}+\delta_{\mu\nu}\,x_0\right)\,{\mathbf y}_\nu^{(1)}\cdot{\mathbf y}_\nu^{(1)} \\
  \label{TGS41b}
  &\stackrel{(\ref{TGS27b})}{=}& 2\left(N-1+\kappa_\mu^{(2)} \right) +x_0\,{\mathbf y}_\mu^{(1)}\cdot{\mathbf y}_\mu^{(1)}
  \;,
\end{eqnarray}
and further
\begin{eqnarray}
\label{TGS42a}
  \sum_{\mu\nu}{\mathbbm J}_{\mu\nu}({\boldsymbol\lambda}^{(0)},\gamma_0)\,
  \left({\mathbf y}_\nu^{(1)}\cdot{\mathbf y}_\nu^{(1)}\right)\,\left({\mathbf y}_\mu^{(1)}\cdot{\mathbf y}_\mu^{(1)}\right)
  &\stackrel{(\ref{TGS41b})}{=}&
  2\sum_{\mu}\left(N-1+\kappa_\mu^{(2)} \right)\,{\mathbf y}_\mu^{(1)}\cdot{\mathbf y}_\mu^{(1)}
  +x_0\sum_{\mu}\left({\mathbf y}_\mu^{(1)}\cdot{\mathbf y}_\mu^{(1)}\right)^2\\
  \label{TGS42b}
   &\stackrel{(\ref{TGS33a},\ref{TGS29})}{=}& -4\,N\,M^{(2)}+x_0\sum_{\mu}\left({\mathbf y}_\mu^{(1)}\cdot{\mathbf y}_\mu^{(1)}\right)^2
   \;.
\end{eqnarray}

%%%%%%%%%%%%%%%%%%%%%%%%%%%%%%%%%%%%%%%%%%%%%%%%%%%%%%%%%%%%%%%%%%%%%%%%%%%%%%%%%%%%%%%%%%%%%%%%%%%%%%%%%%%%%%%%%%%%
\subsubsection{Terms $O(t^4)$:}\label{sec:TGST4}
%%%%%%%%%%%%%%%%%%%%%%%%%%%%%%%%%%%%%%%%%%%%%%%%%%%%%%%%%%%%%%%%%%%%%%%%%%%%%%%%%%%%%%%%%%%%%%%%%%%%%%%%%%%%%%%%%%%%

As in Section \ref{sec:PS4} the fourth order terms of (\ref{PN6}) can be used to determine $\kappa_\mu^{(4)}$ for all $\mu=1,\ldots,N$.
We will not dwell upon the details but rather
consider the fourth order part of the magnetization:
\begin{equation}\label{TGS46}
 M^{(4)}=\sum_\mu x_\mu^{(4)}\stackrel{(\ref{TGS5b})}{=}
 -\sum_\mu\left( {\mathbf y}_\mu^{(1)}\cdot{\mathbf y}_\mu^{(3)}+\frac{1}{8}\left({\mathbf y}_\mu^{(1)}\cdot{\mathbf y}_\mu^{(1)} \right)^2\right)
 \;.
\end{equation}

For the fourth order correction to the energy we obtain:
\begin{eqnarray}
\label{TGS47a}
 E^{(4)} &=& \frac{1}{2}\sum_{\mu\nu}{\mathbbm J}_{\mu\nu}({\mathbf 0},0)\left(
 2\,{\mathbf s}_\mu^{(0)}\cdot{\mathbf s}_\nu^{(4)}+2\,{\mathbf s}_\mu^{(1)}\cdot{\mathbf s}_\nu^{(3)}+{\mathbf s}_\mu^{(2)}\cdot{\mathbf s}_\nu^{(2)}
 \right)\\
 \label{TGS47b}
  &\stackrel{(\ref{TGS5a},\ref{TGS5b})}{=}&
  \frac{1}{2}\sum_{\mu\nu}{\mathbbm J}_{\mu\nu}({\mathbf 0},0)
  \left( -2{\mathbf y}_\nu^{(1)}\cdot{\mathbf y}_\nu^{(3)}-\frac{1}{4}\left({\mathbf y}_\nu^{(1)}\cdot{\mathbf y}_\nu^{(1)} \right)^2
+ 2{\mathbf y}_\mu^{(1)}\cdot{\mathbf y}_\nu^{(3)}+\frac{1}{4}\left({\mathbf y}_\mu^{(1)}\cdot{\mathbf y}_\mu^{(1)} \right)
\left({\mathbf y}_\nu^{(1)}\cdot{\mathbf y}_\nu^{(1)} \right)
 \right)
 \;.
 \end{eqnarray}
Since the bracket in the last equation vanishes for $\mu=\nu$ we may add arbitrary diagonal elements to ${\mathbbm J}_{\mu\nu}({\mathbf 0},0)$
without changing the total value of $E^{(4)}$. In particular, we may choose the homogeneous gauge of the J-matrix thus obtaining:
\begin{eqnarray}
\label{TGS48a}
 E^{(4)} &=&
  \frac{1}{2}\sum_{\mu\nu}J^{(h)}_{\mu\nu}
  \left( -2{\mathbf y}_\nu^{(1)}\cdot{\mathbf y}_\nu^{(3)}-\frac{1}{4}\left({\mathbf y}_\nu^{(1)}\cdot{\mathbf y}_\nu^{(1)} \right)^2
+ 2{\mathbf y}_\mu^{(1)}\cdot{\mathbf y}_\nu^{(3)}+\frac{1}{4}\left({\mathbf y}_\mu^{(1)}\cdot{\mathbf y}_\mu^{(1)} \right)
\left({\mathbf y}_\nu^{(1)}\cdot{\mathbf y}_\nu^{(1)} \right)
 \right)\\
 \nonumber
 &\stackrel{(\ref{S4},\ref{S5},\ref{S1})}{=}& -j \sum_\nu
 \left( {\mathbf y}_\nu^{(1)}\cdot{\mathbf y}_\nu^{(3)}+\frac{1}{8}\left({\mathbf y}_\nu^{(1)}\cdot{\mathbf y}_\nu^{(1)} \right)^2 \right)\\
 \label{TGS48b}
 &&
 +
 \sum_{\mu\nu}\left({\mathbbm J}_{\mu\nu}({\boldsymbol\lambda}^{(0)},\gamma_0)-\gamma_0\,\Xi_{\mu\nu} \right)
 \left({\mathbf y}_\mu^{(1)}\cdot{\mathbf y}_\nu^{(3)}+\frac{1}{8}\left({\mathbf y}_\mu^{(1)}\cdot{\mathbf y}_\mu^{(1)} \right)
 \left({\mathbf y}_\nu^{(1)}\cdot{\mathbf y}_\nu^{(1)} \right)
 \right)\\
  \nonumber
 &\stackrel{(\ref{PN46},\ref{PN22},\ref{S12b},\ref{S2})}{=}&
 j\,M^{(4)}+x_0\,\sum_\nu  {\mathbf y}_\nu^{(1)}\cdot{\mathbf y}_\nu^{(3)}+\gamma_0\,\sum_\nu {\mathbf y}_\nu^{(1)}\cdot{\mathbf y}_\nu^{(3)}+
  \frac{1}{8}\sum_{\mu\nu}{\mathbbm J}_{\mu\nu}({\boldsymbol\lambda}^{(0)},\gamma_0)\,\left({\mathbf y}_\mu^{(1)}\cdot{\mathbf y}_\mu^{(1)} \right)
 \left({\mathbf y}_\nu^{(1)}\cdot{\mathbf y}_\nu^{(1)} \right)\\
  \label{TGS48c}
 &&
 -\frac{\gamma_0}{8}
 \left[
 \left(\sum_\mu {\mathbf y}_\mu^{(1)}\cdot{\mathbf y}_\mu^{(1)} \right)\,\left(\sum_\nu {\mathbf y}_\nu^{(1)}\cdot{\mathbf y}_\nu^{(1)}\right)
 -\sum_\nu \left({\mathbf y}_\nu^{(1)}\cdot{\mathbf y}_\nu^{(1)} \right)^2
 \right]\\
 \nonumber
 &\stackrel{(\ref{PN23})}{=}&
 j M^{(4)}+x_0\sum_\nu  {\mathbf y}_\nu^{(1)}\cdot{\mathbf y}_\nu^{(3)}+\gamma_0\sum_\nu
 \left(   {\mathbf y}_\nu^{(1)}\cdot{\mathbf y}_\nu^{(3)} +\frac{1}{8}\left({\mathbf y}_\nu^{(1)}\cdot{\mathbf y}_\nu^{(1)} \right)^2\right)\\
  \label{TGS48d}
  &&
  +
 \frac{1}{8}\sum_{\mu\nu}{\mathbbm J}_{\mu\nu}({\boldsymbol\lambda}^{(0)},\gamma_0)\left({\mathbf y}_\mu^{(1)}\cdot{\mathbf y}_\mu^{(1)} \right)
\left({\mathbf y}_\nu^{(1)}\cdot{\mathbf y}_\nu^{(1)} \right) -\frac{\gamma_0}{2}M^{(2)2}\\
&\stackrel{(\ref{TGS42b},\ref{S12a})}{=}&j\, M^{(4)}+(x_0+\gamma_0)\,\sum_\nu
\left(  {\mathbf y}_\nu^{(1)}\cdot{\mathbf y}_\nu^{(3)} +\frac{1}{8}\left({\mathbf y}_\nu^{(1)}\cdot{\mathbf y}_\nu^{(1)} \right)^2\right)
-\frac{N}{2}\,M^{(2)} +\frac{j-j_{min}^{(h)}}{2N}\,M^{(2)2}\\
\label{TGS48e}
&\stackrel{(\ref{TGS46},\ref{S12a})}{=}&\left(j-j_{min}^{(h)}\right)\,M^{(4)}-\frac{N}{2}\,M^{(2)} +\frac{j-j_{min}^{(h)}}{2N}\,M^{(2)2}
\;.
\end{eqnarray}

%%%%%%%%%%%%%%%%%%%%%%%%%%%%%%%%%%%%%%%%%%%%%%%%%%%%%%%%%%%%%%%%%%%%%%%%%%%%%%%%%%%%%%%%%%%%%%%%%%%%%%%%%%%%%%%%%%%%%%%%%%%%%%%%%%%%%%%%%%
\subsubsection{Saturation susceptibility}\label{sec:TGSSS}
%%%%%%%%%%%%%%%%%%%%%%%%%%%%%%%%%%%%%%%%%%%%%%%%%%%%%%%%%%%%%%%%%%%%%%%%%%%%%%%%%%%%%%%%%%%%%%%%%%%%%%%%%%%%%%%%%%%%%%%%%%%%%%%%%%%%%%%%%%

Analogously to the results of Section \ref{sec:SS} we obtain for the leading coefficient
of the $t$-series for the magnetic field
\begin{equation}\label{TGSSS2}
 B_{sat}\equiv \lim_{t\rightarrow 0}B(t)=\frac{E^{(2)}}{M^{(2)}}\stackrel{(\ref{TGS30f})}{=}
  \frac{\left( j-j_{min}^{(h)} \right)\,M^{(2)}}{M^{(2)}}=j-j_{min}^{(h)}
  \;.
\end{equation}
This result is in accordance with (\ref{HZS1}).

Similarly, we reconsider the series representation of the susceptibility
\begin{eqnarray}
\label{TGSSS3a}
  \chi(t)&=& \frac{\partial M/\partial t}{\partial B/\partial t} \\
  &\stackrel{(\ref{SS1c})}{=}&\frac{2\, M^{(2)}\, t+ 4\, M^{(4)\,}t^3+\ldots}{\frac{4t}{M^{(2)2}}\left(E^{(4)}M^{(2)}-E^{(2)}M^{(4)} \right)}
  \;.
\end{eqnarray}
and the corresponding saturation susceptibility
\begin{eqnarray}\label{TGSSS4a}
 \chi_0&\equiv& \lim_{t\rightarrow 0}\chi(t) =\frac{M^{(2)3}}{2\left(E^{(4)}M^{(2)}-E^{(2)}M^{(4)} \right)}\\
 \label{TGSSS4b}
 &=&\frac{M^{(2)3}}{2\left( B_{sat}M^{(4)}+\frac{B_{sat}M^{(2)2}}{2N}-\frac{N}{2}M^{(2)} \right)\,
M^{(2)}-2\, B_{sat}\,M^{(2)}\,M^{(4)}}\\
 \label{TGSSS4c}
 &=& \frac{N}{B_{sat}-\frac{N^2}{M^{(2)}}}\\
 \label{TGSSS4d}
 &\stackrel{(\ref{TGS38})}{=}& \frac{N}{B_{sat}+\frac{2 N^2}{\alpha_{11}^2+2 \alpha_{12}^2+\alpha_{22}^2}}
 \equiv \frac{N}{B_{sat}+N\,k^2}
 \;.
\end{eqnarray}
In Eq.~(\ref{TGSSS4b}) we have inserted the
previous results for $E^{(2)}$ and $E^{(4)}$, see (\ref{TGS30f}) and (\ref{TGS48e}), as well as (\ref{TGSSS2}).
In the last equation (\ref{TGSSS4d}) we have written the saturation susceptibility in a form analogous to the coplanar case (\ref{SS4d}).

%%%%%%%%%%%%%%%%%%%%%%%%%%%%%%%%%%%%%%%%%%%%%%%%%%%%%%%%%%%%%%%%%%%%%%%%%%%%%%%%%%%%%%%%%%%%%%%%%%%%%%%%%%%%%%%%%%%%%%%%%%%%%%%%%%%%%%%%%%
\section{Examples}\label{sec:E}
%%%%%%%%%%%%%%%%%%%%%%%%%%%%%%%%%%%%%%%%%%%%%%%%%%%%%%%%%%%%%%%%%%%%%%%%%%%%%%%%%%%%%%%%%%%%%%%%%%%%%%%%%%%%%%%%%%%%%%%%%%%%%%%%%%%%%%%%%%
Examples for parabolic systems including the odd regular polygons can also be found in \cite{S17d}.
We will add an example in Section \ref{sec:IT} that is only ``locally parabolic", i.~e., for
a certain interval of magnetization $M_1\le M\le N$ in order to support our proposal to weaken the
pertinent definition.

%%%%%%%%%%%%%%%%%%%%%%%%%%%%%%%%%%%%%%%%%%%%%%%%%%%%%%%%%%%%%%%%%%%%%%%%%%%%%%%%%%%%%%%%%%%%%%%%%%%%%%%%%%%%%%%%%%%%%%%%%%%%%%%%%%%%%%%%%%
\subsection{Irregular tetrahedron}\label{sec:IT}
%%%%%%%%%%%%%%%%%%%%%%%%%%%%%%%%%%%%%%%%%%%%%%%%%%%%%%%%%%%%%%%%%%%%%%%%%%%%%%%%%%%%%%%%%%%%%%%%%%%%%%%%%%%%%%%%%%%%%%%%%%%%%%%%%%%%%%%%%%

\begin{figure}[t]
\centering
\includegraphics[width=0.50\linewidth]{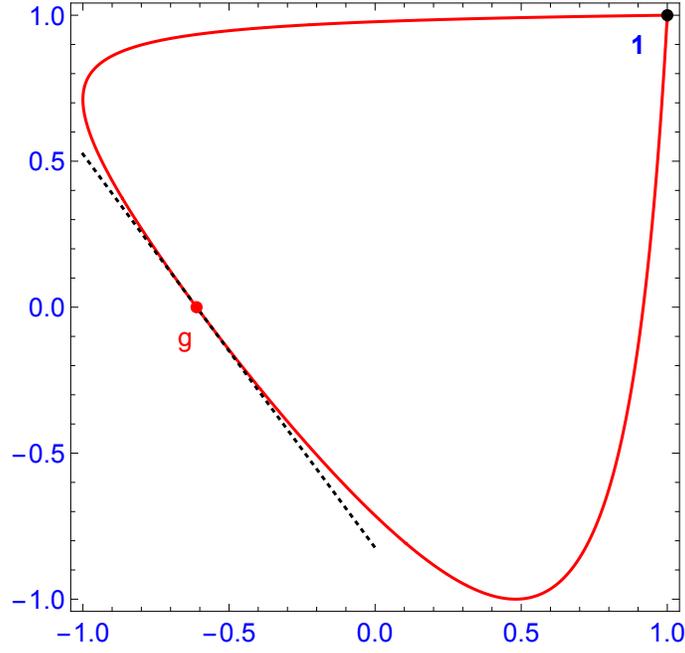}
\caption{The convex domain ${\mathcal G}$ of $u,v$-values such that the Gram matrix $G(u,v)$ according to (\ref{IT4}) will be positively semi-definite.
The (black) point corresponding to the ferromagnetic ground state $\uparrow\uparrow\uparrow\uparrow$ is marked by ``${\mathbf 1}$",
analogously the (red) point corresponding to the coplanar ground state with minimal magnetization (or energy) by ``$g$". The tangent
at $g$ (black dotted line) has been calculated by means of (\ref{IT5a}) or (\ref{IT5b}).
}
\label{FIGGIT}
\end{figure}

As an example of a parabolic system we consider a tetrahedron ($N=4$) with six coupling coefficients
which are chosen so that the example fulfils its purpose mentioned above. The homogeneously gauged J-matrix is taken as
\begin{equation}\label{IT1}
 J^{(h)}=\frac{1}{262}\left(
\begin{array}{cccc}
 235 & -141 & -291 & 459 \\
 -141 & -125 & 489 & 39 \\
 -291 & 489 & 95 & -31 \\
 459 & 39 & -31 & -205 \\
\end{array}
\right)
\;.
\end{equation}
Its eigenvalues are
\begin{equation}\label{IT2}
j_1=3,\quad j_{min}^{(h)}=-2 \mbox{ (twofold degenerate) and }j=1
\;,
\end{equation}
which entails
\begin{equation}\label{IT3}
 B_{sat}=j-j_{min}^{(h)}=3, \quad \gamma_0=\frac{j_{min}^{(h)}-j}{N}=-\frac{3}{4}
 \;.
\end{equation}
The ADE (\ref{T10}) of the corresponding J-matrix with uniform coupling ${\mathbbm J}\left({\boldsymbol\lambda}^{(0)},\gamma_0\right)$
has solutions depending on two parameters $u,v$ such that the corresponding Gram matrix reads
\begin{equation}\label{IT4}
G(u,v)=\left(\begin{array}{cccc}
 1 & \frac{1}{135} (91 u+44) & \frac{1}{65} (21 v+44) & \frac{1}{15} (13 u+9 v-7)
   \\
 \frac{1}{135} (91 u+44) & 1 & \frac{1}{117} (91 u+63 v-37) & v \\
 \frac{1}{65} (21 v+44) & \frac{1}{117} (91 u+63 v-37) & 1 & u \\
 \frac{1}{15} (13 u+9 v-7) & v & u & 1 \\
\end{array}
\right)
\;.
\end{equation}
The convex set ${\mathcal G}$ in the $(u,v)$-plane corresponding to those points where $G(u,v)\ge 0$ and hence to physical ground states
is shown in Figure \ref{FIGGIT}. Its boundary corresponds to coplanar ground states except the point ${\mathbf 1}$ representing
the ferromagnetic ground state $\uparrow\uparrow\uparrow\uparrow$. The points in the interior of ${\mathcal G}$ correspond to
three-dimensional states of the form (\ref{P1}).

\begin{figure}[t]
\centering
\includegraphics[width=0.30\linewidth]{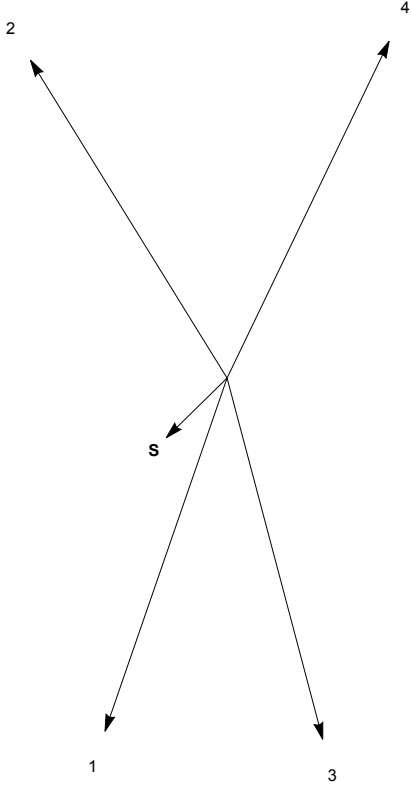}
\caption{The absolute coplanar ground state ${\mathbf h}$ of the irregular tetrahedron determined numerically and the total spin ${\mathbf S}$
of length $M_0=0.22834$.
}
\label{FIGGSIT}
\end{figure}

\begin{figure}[t]
\centering
\includegraphics[width=0.50\linewidth]{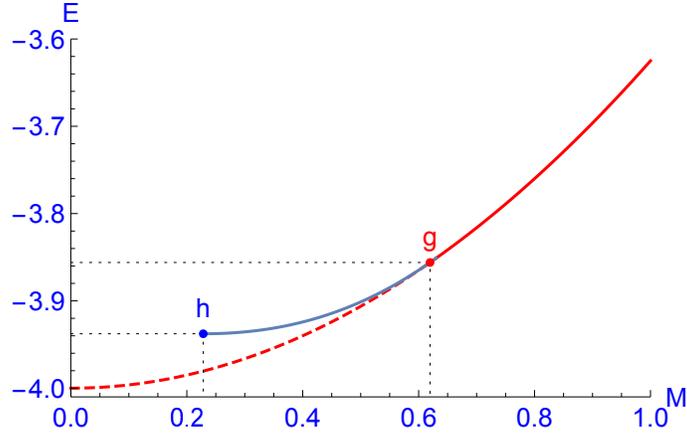}
\caption{The minimal energy $E$ as a function of magnetization $M$ for relative ground states of the irregular tetrahedron and $M_0\le M\le 1$.
The solid red parabola represents a subset of ground states given by the points of ${\mathcal G}$ and satisfying the equation (\ref{IT6}).
The ground state of ${\mathcal G}$ with minimal magnetization is marked by ``$g$" as in Figure \ref{FIGGIT}.
The dashed red parabola represents the continuation of (\ref{IT6}) to lower values of $M$. The blue curve is a fit
of $E(M)$ for $75$ numerically determined coplanar ground states including the absolute ground state marked by ``$h$".
Obviously the numerically determined ground states have an energy above the parabola (\ref{IT6}).
}
\label{FIGGSIT2}
\end{figure}
It will be instructive to calculate the energy (without uniform coupling) $E(u,v)$ and the squared magnetization $M^2(u,v)$
for the ground states corresponding to the points of ${\mathcal G}$. The result is
\begin{eqnarray}
\label{IT5a}
 E(u,v)&=& \frac{1}{2}\mbox{Tr}\left( G(u,v) J^{(h)}\right)=\frac{2}{585} (728 u+540 v-683), \\
 \label{IT5b}
 M^2(u,v) &=& \sum_{\mu\nu}G_{\mu\nu}(u,v) =\frac{16 (728 u+540 v+487)}{1755}
 \;.
\end{eqnarray}
These two functions satisfy the linear relation
\begin{equation}\label{IT6}
 E(u,v)=-4 +\frac{3}{8}\,M^2(u,v)=  \frac{1}{2}\, N\,j_{min}^{(h)}+\frac{j-j_{min}^{(h)}}{2\,N}\, M^2(u,v)
 \;,
\end{equation}
showing that the present example of the irregular tetrahedron is parabolic in the sense of Section \ref{sec:P} and, in particular, has a constant susceptibility
$\chi= \frac{4}{3}$ in the domain ${\mathcal G}$.

The state $g$ with the lowest magnetization $M_1=0.619623$ (or lowest energy $E_1=-3.85603$)
among the states corresponding to ${\mathcal G}$ is not the absolute ground state. We have numerically determined
the absolute ground state with $M_0=0.22834$ and $E_0=-3.93768$, see Figure \ref{FIGGSIT},  and the relative (coplanar) ground states for $M_0<M<M_1$,
see Figure \ref{FIGGSIT2}. Obviously, the energies of these relative ground states are above the parabola (\ref{IT6}) and
hence the irregular tetrahedron is an example of a parabolic system in the sense of Section \ref{sec:P}
that is not parabolic for all physical possible values of the magnetization.

%%%%%%%%%%%%%%%%%%%%%%%%%%%%%%%%%%%%%%%%%%%%%%%%%%%%%%%%%%%%%%%%%%%%%%%%%%%%%%%%%%%%%%%%%%%%%%%%%%%%%%%%%%%%%%%%%%%%%%%%%%%%%%%%%%%%%%%%%%
\subsection{Isosceles triangle}\label{sec:ET}
%%%%%%%%%%%%%%%%%%%%%%%%%%%%%%%%%%%%%%%%%%%%%%%%%%%%%%%%%%%%%%%%%%%%%%%%%%%%%%%%%%%%%%%%%%%%%%%%%%%%%%%%%%%%%%%%%%%%%%%%%%%%%%%%%%%%%%%%%%

\begin{figure}[t]
\centering
\includegraphics[width=0.70\linewidth]{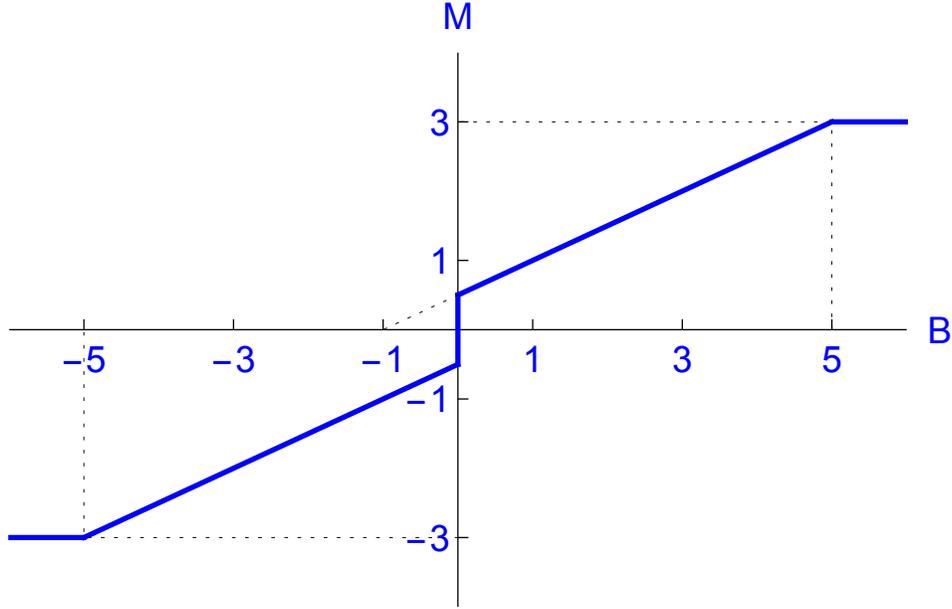}
\caption{The magnetization $M$ as a function of the magnetic field $B$ for the isosceles spin triangle considered in this section.
We have extended the domain of $B$ to include also negative values and plotted $M(B)$ as an odd function.
The magnetization $M(B)$ is the linear function $M(B)=1/2+B/2$ in the interval $0<B<B_{sat}=5$ in accordance with the result $\chi=1/2$
obtained for the analytical domain $0<B<B_{sat}$, see (\ref{ET10}).
}
\label{FIGBM2}
\end{figure}

For the non-parabolic case we first we consider a relatively simple example where all quantities can be analytically calculated. This
will be the AF triangle ($N=3$) with coupling coefficients $J_{12}=J_{13}=1$ and $J_{23}=2$. The corresponding
homogeneously gauged $J$-matrix has the form
\begin{equation}\label{ET1}
J^{(h)}=\left(
\begin{array}{ccc}
 \frac{2}{3} & 1 & 1 \\
 1 & -\frac{1}{3} & 2 \\
 1 & 2 & -\frac{1}{3} \\
\end{array}
\right)
\;,
\end{equation}
and its eigenvalues are
\begin{equation}\label{ET2}
j=\frac{8}{3},\quad j_{min}^{(h)}=-\frac{7}{3},\quad j_3=-\frac{1}{3}
\;,
\end{equation}
with (normalized) eigenvectors
\begin{equation}\label{ET2a}
{\mathbf e}_1=\frac{1}{\sqrt{3}}
\left(\begin{array}{r}
        1 \\
        1 \\
        1
      \end{array}
\right)
,\quad {\mathbf e}_2={\boldsymbol\xi}
=\frac{1}{\sqrt{2}}
\left(\begin{array}{r}
        0 \\
        1 \\
        -1
      \end{array}
\right)
,\quad {\mathbf e}_3=\frac{1}{\sqrt{6}}
\left(\begin{array}{r}
        -2 \\
        1 \\
        1
      \end{array}
\right)
\;.
\end{equation}
From this we calculate the saturation field
\begin{equation}\label{ET3}
 B_{sat}\stackrel{(\ref{SS2})}{=}j-j_{min}^{(h)}=5
\end{equation}
and the critical uniform coupling parameter
\begin{equation}\label{ET4}
 \gamma_0\stackrel{(\ref{S12b})}{=}\frac{j_{min}^{(h)}-j}{N}=-\frac{5}{3}
 \;.
\end{equation}
Since the ground state problem for the general triangle has been completely solved in \cite{S17c}
it will suffice to give the following results without detailed derivation:
\begin{equation}\label{ET5}
 {\mathbbm J}({\boldsymbol\lambda}(t),\gamma_0+t^2)=
\left(
\begin{array}{ccc}
 \frac{2}{3 t^2+1}-\frac{4}{3} & t^2-\frac{2}{3} & t^2-\frac{2}{3} \\
 t^2-\frac{2}{3} & \frac{1}{-3 t^2-1}+\frac{2}{3} & t^2+\frac{1}{3} \\
 t^2-\frac{2}{3} & t^2+\frac{1}{3} & \frac{1}{-3 t^2-1}+\frac{2}{3} \\
\end{array}
\right)
\;,
\end{equation}
\begin{eqnarray}
\label{ET6a}
  {\mathbf s}_1(t) &=& {1\choose 0}\;, \\
  \label{ET6b}
  {\mathbf s}_2(t) &=&{\frac{2-3 t^2}{6 t^2+2} \choose \frac{3 t \sqrt{3 t^2+4}}{6 t^2+2}}=
  {1\choose 0} + t\,{0\choose 3} + t^2\,{-\frac{9}{2}\choose 0}+ t^3\,{0\choose -\frac{63}{8}}+ t^4\,{\frac{27}{2}\choose 0} +\ldots \\
  \label{ET6c}
  {\mathbf s}_3(t) &=& {\frac{2-3 t^2}{6 t^2+2}\choose -\frac{3 t \sqrt{3 t^2+4}}{6 t^2+2}}=
  {1\choose 0} + t\,{0\choose -3} + t^2\,{-\frac{9}{2}\choose 0}+ t^3\,{0\choose \frac{63}{8}}+ t^4\,{\frac{27}{2}\choose 0} +\ldots
  \;.
\end{eqnarray}
For $t=t_0\equiv \sqrt{\frac{5}{3}}$ the absolute ground state
\begin{equation}\label{ET6d}
{\mathbf s}_1(t_0)=\left(
\begin{array}{c}
 1 \\
 0 \\
\end{array}
\right),
\quad
{\mathbf s}_2(t_0)=
\left(
\begin{array}{c}
 -\frac{1}{4} \\
 \frac{\sqrt{15}}{4} \\
\end{array}
\right),
\quad
{\mathbf s}_3(t_0)=
\left(
\begin{array}{c}
 -\frac{1}{4} \\
 -\frac{\sqrt{15}}{4} \\
\end{array}
\right),
\end{equation}
with a residual magnetization of $M=1/2$ will be assumed.
Further,
\begin{equation}\label{ET7}
  M(t)=\frac{3}{3 t^2+1}=3-9 t^2+27 t^4+\ldots
  \;,
\end{equation}
\begin{eqnarray}
\label{ET7a}
 \kappa_1(t) &=& \frac{\left(2-3 t^2\right)^2}{9 t^2+3}=\frac{4}{3}-8 t^2+27 t^4+\ldots \;,\\
 \label{ET7b}
 \kappa_2(t)=\kappa_3(t) &=& \frac{1}{3}+t^2
 \;.
\end{eqnarray}
The latter yields the  minimal eigenvalue $x(t)$ of  ${\mathbbm J}({\boldsymbol\lambda}(t),\gamma_0+t^2)$
by summation over $\mu=1,2,3$:
\begin{equation}\label{ET7c}
 x(t)=-\frac{1}{N}\sum_\mu \kappa_\mu(t)=-\frac{9 t^4+2}{9 t^2+3}=-\frac{2}{3}+2 t^2-9 t^4+\ldots
 \;.
\end{equation}
For the total energy (without uniform coupling) we obtain
\begin{equation}\label{ET8}
  E(t)=\frac{-18 t^4-21 t^2+4}{\left(3 t^2+1\right)^2}=4-45 t^2+216 t^4+\ldots
  \;,
\end{equation}
further
\begin{equation}\label{ET9}
  B(t)=\frac{\partial E(t)/\partial t}{\partial M(t)/\partial t}=\frac{5-3 t^2}{1+3 t^2}=5-18 t^2+54 t^4+\ldots
  \;,
\end{equation}
and finally
\begin{equation}\label{ET10}
\chi(t)=\frac{\partial M(t)/\partial t}{\partial B(t)/\partial t}=\frac{1}{2}
\;,
\end{equation}
that turns out to be constant for $0<t<\sqrt{\frac{5}{3}}$, see Figure \ref{FIGBM2}.
Although the minimal energy is a quadratic function of the magnetization, $E(t)=-2-M(t)+M^2(t)$,
the system is not parabolic in the sense of Section \ref{sec:P} since its susceptibility
is not given by $\frac{N}{B_{sat}}=\frac{3}{5}$ as it should be for parabolic systems according to (\ref{P8}).

Thus the physical quantities and their expansions into $t$-series are completely known for the considered isosceles triangle
and one may directly check the results of Section \ref{sec:CGS}. We will confine ourselves to a few significant cases.
First we compare the eigenvector ${\boldsymbol\xi}=\frac{1}{\sqrt{2}}(0,1,-1)^\top$ of $J^{(h)}$ with the vector
$x^{(1)}=(0,3,-3)^\top$ of linear ground state corrections and conclude
\begin{equation}\label{ET11}
 X=3\,\sqrt{2}
 \;,
\end{equation}
by means of (\ref{PN23}).
Let ${\boldsymbol\xi}^2=\frac{1}{2}(0,1,1)^\top)$ be the vector of squared components $\xi_\mu^2,\,\mu=1,2,3$.
To check (\ref{PN37}) we note that the operator $K$ defined in (\ref{PN24}) will be of the form
$K=|{\mathbf k}\rangle\langle {\mathbf k}|$ with ${\mathbf k}=\frac{1}{\sqrt{3}}(-2,1,1)^\top=\sqrt{2}\,{\mathbf e}_3$. It follows that
\begin{equation}\label{ET12}
 k^2=\sum_{\mu\nu}K_{\mu\nu}\,\xi_\nu^2\,\xi_\mu^2= \left| \langle{\mathbf k}|{\boldsymbol\xi}^2\rangle\right|^2
 =\left| \frac{1}{\sqrt{3}}
\left(\begin{array}{r}
       -2 \\
        1 \\
        1
      \end{array}
\right)\cdot\frac{1}{2}
\left(\begin{array}{r}
        0 \\
        1 \\
        1
      \end{array}
\right)\right|^2
=\frac{1}{3}
\;.
\end{equation}
Hence
\begin{equation}\label{ET13}
X \stackrel{(\ref{ET11})}{=}3\,\sqrt{2}=\frac{\sqrt{2\times 3}}{\frac{1}{\sqrt{3}}}\stackrel{(\ref{ET12})}{=}\frac{\sqrt{2\,N}}{k}
\;,
\end{equation}
thereby confirming (\ref{PN37}).
Finally, we will check Eq.~ (\ref{SS4d}):
\begin{equation}\label{ET14}
\chi\stackrel{(\ref{ET10})}{=}\frac{1}{2}=\frac{3}{5+3\times\frac{1}{3}}\stackrel{(\ref{ET3},\ref{ET12})}{=}\frac{N}{B_{sat}+N\,k^2}
\;.
\end{equation}

%%%%%%%%%%%%%%%%%%%%%%%%%%%%%%%%%%%%%%%%%%%%%%%%%%%%%%%%%%%%%%%%%%%%%%%%%%%%%%%%%%%%%%%%%%%%%%%%%%%%%%%%%%%%%%%%%%%%%%%%%%%%%%%%%%%%%%%%%%
\subsection{Almost regular Cube}\label{sec:EC}
%%%%%%%%%%%%%%%%%%%%%%%%%%%%%%%%%%%%%%%%%%%%%%%%%%%%%%%%%%%%%%%%%%%%%%%%%%%%%%%%%%%%%%%%%%%%%%%%%%%%%%%%%%%%%%%%%%%%%%%%%%%%%%%%%%%%%%%%%%

\begin{figure}[t]
\centering
\includegraphics[width=0.50\linewidth]{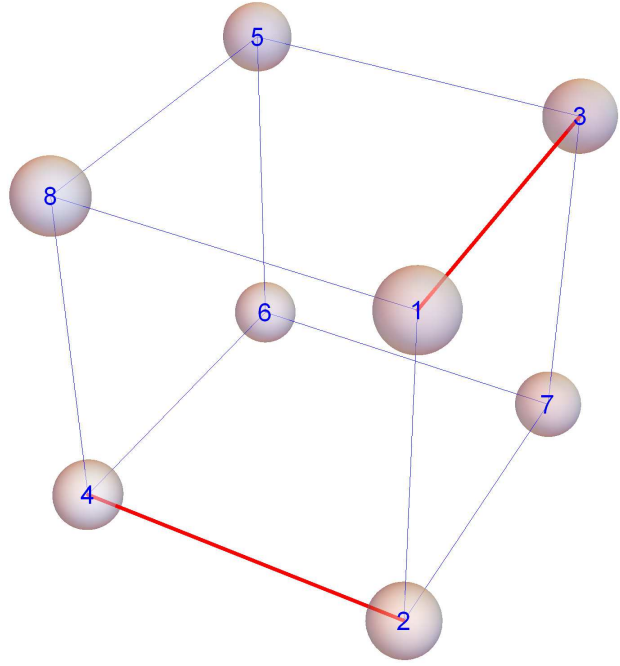}
\caption{The almost regular cube with ten AF coupling paramaters $J_{\mu\nu}=+1$ (blue lines) and two ferromagnetic
bonds $J_{24}=J_{13}=-1$ (red lines).
}
\label{FIGCUBE}
\end{figure}

\begin{figure}[t]
\centering
\includegraphics[width=0.50\linewidth]{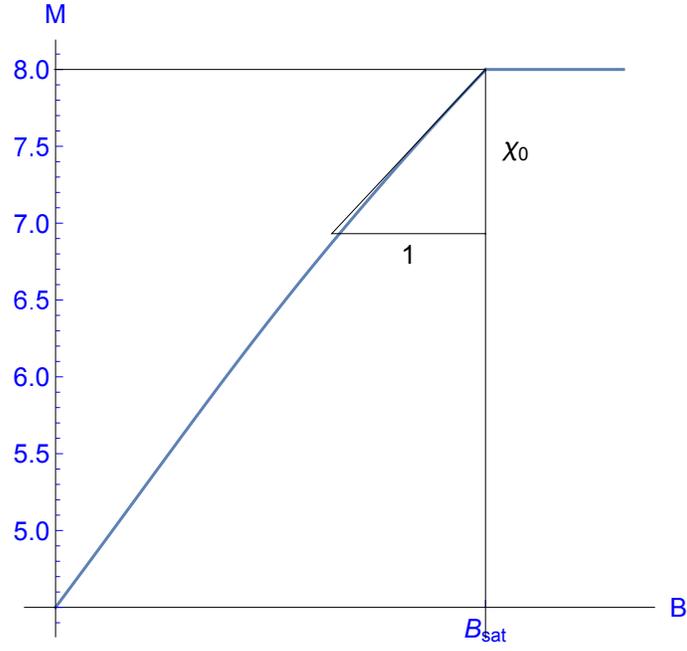}
\caption{The numerically determined magnetization curve $M(B)$ of the almost regular cube in the vicinity of the saturation point
and the slope triangle with the analytically determined slope $\chi_0$ according to (\ref{EC8}).
}
\label{FIGBM3}
\end{figure}

We consider a cube ($N=8$) with AF coupling  $J_{\mu\nu}=+1$  except two ferromagnetic bonds $J_{24}=J_{13}=-1$, see
Figure \ref{FIGCUBE}. We will analytically calculate the saturation susceptibility and check it by numerical calculations.
This example has also be considered in \cite{SF20} with a general ferromagnetic bond strength.
Its  homogeneously gauged $J$-matrix  has the form
\begin{equation}\label{EC1}
  J^{(h)}=\left(
\begin{array}{cccccccc}
 1 & 1 & -1 & 0 & 0 & 0 & 0 & 1 \\
 1 & 1 & 0 & -1 & 0 & 0 & 1 & 0 \\
 -1 & 0 & 1 & 0 & 1 & 0 & 1 & 0 \\
 0 & -1 & 0 & 1 & 0 & 1 & 0 & 1 \\
 0 & 0 & 1 & 0 & -1 & 1 & 0 & 1 \\
 0 & 0 & 0 & 1 & 1 & -1 & 1 & 0 \\
 0 & 1 & 1 & 0 & 0 & 1 & -1 & 0 \\
 1 & 0 & 0 & 1 & 1 & 0 & 0 & -1 \\
\end{array}
\right)\;.
\end{equation}
The corresponding characteristic polynomial reads
\begin{equation}\label{EC2}
  p(x)=(x-2) x (x+2) \left(x^2-2 x-2\right) \left(x^3+2 x^2-6 x-8\right)
  \;,
\end{equation}
which leads to the two prominent eigenvalues
\begin{equation}\label{EC3}
 j=2,\quad j_{min}^{(h)}=\text{Root}\left[x^3+2 x^2-6 x-8,1\right]=-3.10278\ldots
 \;.
\end{equation}
with corresponding (not normalized) eigenvectors ${\mathbf 1}$ and
\begin{eqnarray}\nonumber
 {\boldsymbol\xi}&=&\left(\text{Root}\left[4 x^3-7 x-1,2\right],-\text{Root}\left[4 x^3-7
   x-1,2\right],\text{Root}\left[2 x^3+4 x^2-x-1,3\right],\right.\\
   \label{EC4a}
   && \left.-\text{Root}\left[2 x^3+4 x^2-x-1,3\right]  ,\text{Root}\left[4 x^3+4 x^2-3
   x-1,1\right],-\text{Root}\left[4 x^3+4 x^2-3 x-1,1\right],-1,1\right)^\top\\
    \label{EC4b}
   &\approx&
  (-0.144584, 0.144584, 0.551388, -0.551388, -1.4068, 1.4068, -1, 1)^\top
 \;.
\end{eqnarray}
Here we have adopted the notation $\text{Root}\left[p(x),n\right]$ for the $n$th root of the
polynomial $p(x)$ analogous to the similar MATHEMATICA$^\circledR$ command.
In passing we note the special form of $ {\boldsymbol\xi}$ with alternating components
due to the reflectional symmetry of the almost regular cube, see \cite{SF20} for details.
We conclude
\begin{equation}\label{EC5}
 B_{sat}\stackrel{(\ref{SS2})}{=}j-j_{min}^{(h)}=5.10278\ldots,
\end{equation}
and
\begin{equation}\label{EC6}
\gamma_0\stackrel{(\ref{S12b})}{=}\frac{j_{min}^{(h)}-j}{N}=-0.637847\ldots
\;.
\end{equation}
From this we will obtain the matrix $K$, see (\ref{PN24}), and its expectation value $k^2$ according to (\ref{PN36}),
taking into account that (\ref{EC4a}) was not normalized. The exact result reads
\begin{equation}\label{EC7}
 k^2=\text{Root}\left[199712 x^3-46136 x^2-14439 x+3113,3\right]=0.298158\ldots
 \;,
\end{equation}
and yields the saturation susceptibility
\begin{equation}\label{EC8}
 \chi_0\stackrel{(\ref{SS4d})}{=}  \frac{N}{B_{sat}+N k^2}=632\, \text{Root}\left[41295512 x^3+40484 x^2-778 x+1,2\right]
 =1.06837\ldots
 \;.
\end{equation}
To check this exact result we have numerically calculated $137$  ground states for $\gamma_0<\gamma<-1/2$
and fitted the function $E(M)$ by an even  polynomial of degree $12$. This yields a numerical approximation
of $M(B)$, see Figure \ref{FIGBM3}, with a slope $1.037$ at the saturation point in approximate accordance with (\ref{EC8}).

%%%%%%%%%%%%%%%%%%%%%%%%%%%%%%%%%%%%%%%%%%%%%%%%%%%%%%%%%%%%%%%%%%%%%%%%%%%%%%%%%%%%%%%%%%%%%%%%%%%%%%%%%%%%%%%%%%%%%%%%%%%%%%%%%%%%%%%%%%
\subsection{Irregular octahedron}\label{sec:IH}
%%%%%%%%%%%%%%%%%%%%%%%%%%%%%%%%%%%%%%%%%%%%%%%%%%%%%%%%%%%%%%%%%%%%%%%%%%%%%%%%%%%%%%%%%%%%%%%%%%%%%%%%%%%%%%%%%%%%%%%%%%%%%%%%%%%%%%%%%%

\begin{figure}[t]
\centering
\includegraphics[width=0.10\linewidth]{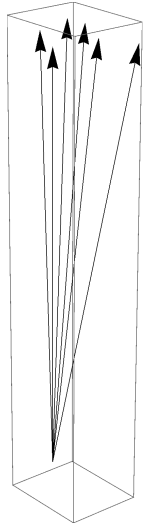}
\caption{The ground state of the irregular octahedron for the value $\gamma=\gamma_0+0.0002$ of the auxiliary uniform coupling.
}
\label{FIGGS4}
\end{figure}

\begin{figure}[t]
\centering
\includegraphics[width=0.70\linewidth]{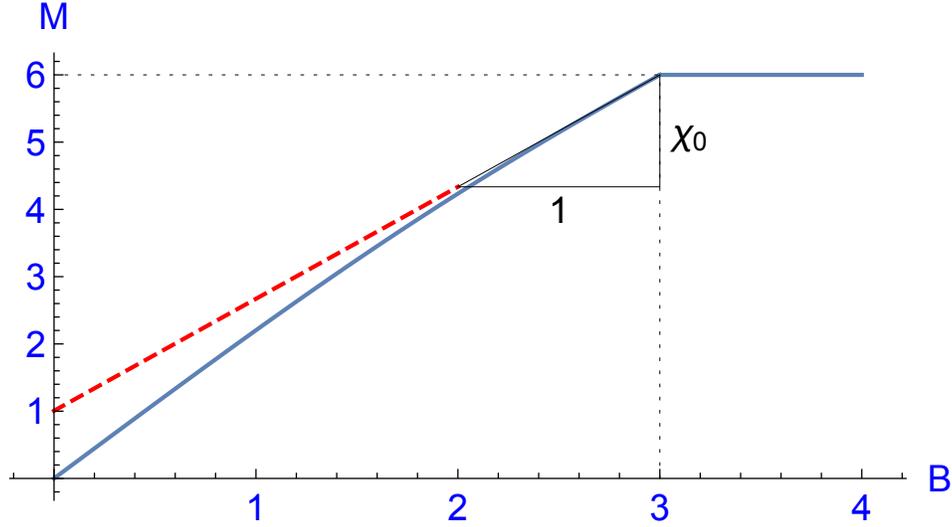}
\caption{The numerically determined magnetization curve $M(B)$ of the irregular octahedron for $B=0,\ldots,4$
and the slope triangle with the analytically determined slope $\chi_0$ of $M(B)$ at the saturation point according to (\ref{IH8}).
}
\label{FIGEX4}
\end{figure}

In order to construct an example of a non-parabolic system with three-dimensional ground states
we consider three vectors
\begin{equation}\label{IH1}
 {\mathbf a}=(1,-2,-1,0,1,1)^\top,\quad {\mathbf b}=(-2, 1, -1, 1, 2, -1)^\top,
 \quad {\mathbf c}=(-4, -2, 3, 0, 0, 3)^\top
 \;,
\end{equation}
orthogonal to
${\mathbf 1}=(1,1,1,1,1,1)^\top $ and mutually orthogonal,
the two-dimensional subspace of ${\mathbbm R}^6$ spanned by ${\mathbf a}$ and ${\mathbf b}$
and the projector $P$ onto this subspace.
Further consider the one-dimensional projectors
$F=\frac{1}{6}\,|{\mathbf 1}\rangle\langle {\mathbf 1}|$ and $Q=\frac{1}{38}|{\mathbf c}\rangle\langle {\mathbf c}|$.
and define the homogeneously gauged $J$-matrix of the ``irregular octahedron" ($N=6$)  by
\begin{equation}\label{IH2}
 J^{(h)}=-2 P + F + 3 Q=
 \frac{1}{2622}
 \left(
\begin{array}{cccccc}
 1697 & 3803 & -2389 & 1235 & 1577 & -3301 \\
 3803 & -1471 & -1831 & 209 & 1235 & 677 \\
 -2389 & -1831 & 932 & 1007 & 2375 & 2528 \\
 1235 & 209 & 1007 & -19 & -589 & 779 \\
 1577 & 1235 & 2375 & -589 & -2527 & 551 \\
 -3301 & 677 & 2528 & 779 & 551 & 1388 \\
\end{array}
\right)
 \;,
\end{equation}
such that its eigenvalues are
\begin{equation}\label{IH3}
 j_1=3,\quad j_{min}^{(h)}=-2\; (\mbox{twofold degenerate}),\quad j=1,\quad j_2=0\; (\mbox{twofold degenerate}).
\end{equation}
It follows that
\begin{equation}\label{IH4}
  B_{sat}\stackrel{(\ref{TGSSS2})}{=}j-j_{min}^{(h)}=3,
  \quad \mbox{and}\quad
  \gamma_0 \stackrel{(\ref{S12b})}{=}\frac{j_{min}^{(h)}-j}{N}=-\frac{1}{2}
  \;.
\end{equation}
Hence the minimal eigenvalue $x_0\stackrel{(\ref{S12a})}{=}j+(N-1)\gamma_0=-\frac{3}{2}$
of ${\mathbbm J}\left({\boldsymbol\lambda}^{(0)},\gamma_0\right)$ is threefold
degenerate and the corresponding eigenspace ${\mathcal E}_0$ is spanned by the vectors ${\mathbf a},\,{\mathbf b},\,{\mathbf 1}$.
We choose ${\boldsymbol\xi}^{(1)}=\frac{1}{2\sqrt{2}}\,{\mathbf a}$ and ${\boldsymbol\xi}^{(2)}=\frac{1}{3\sqrt{2}}\,{\mathbf b}$
as an orthonormal basis in ${\mathcal E}_1$.
The example is chosen such that the ADE for the subspace ${\mathcal E}_0$ has only one solution corresponding to the ferromagnetic
ground state ${\mathbf 1}$ and hence the present system is non-parabolic and admits three-dimensional ground states.

It is straight forward to calculate the matrix $K={\mathbbm J}\left({\boldsymbol\lambda}^{(0)},\gamma_0\right)-x_0\,{\mathbbm 1}$
and the $k_{ij}$ according to Eq.~(\ref{TGS36}):
\begin{eqnarray}
\label{IH5a}
  k_{11} &=& \frac{20555}{41952}\;, \\
  \label{IH5b}
  k_{12} &=&\frac{5423}{13984 \sqrt{23}}\;, \\
  \label{IH5c}
  k_{13} &=& -\frac{143237}{964896}\;, \\
  \label{IH5d}
  k_{23} &=&-\frac{109153}{321632 \sqrt{23}}\;, \\
  \label{IH5e}
  k_{33} &=&\frac{5812139}{22192608}
\;.
\end{eqnarray}
Using computer-algebraic software the unique solution of the corresponding system of polynomial equations (\ref{TGS37a}) - (\ref{TGS37d}) together with
$\alpha_{12}=\alpha_{21}$ and ${\boldsymbol\alpha}>0$ can be obtained as
\begin{eqnarray}
\label{IH6a}
  \alpha_{11} &=& \frac{1}{48} \sqrt{\frac{9423194077641+451984247 \sqrt{110485905}}{149703945}}\approx 6.41047\;,\\
   \alpha_{12} &=& \frac{4433}{48} \sqrt{\frac{23 \left(11649-\sqrt{110485905}\right)}{149703945}}\approx 1.22104\;, \\
   \alpha_{22} &=& \frac{1}{48} \sqrt{\frac{23 \left(971333547423+19651489
   \sqrt{110485905}\right)}{149703945}}\approx  8.86256
   \;.
\end{eqnarray}
This yields
\begin{eqnarray}\label{IH7a}
M^{(2)}&\stackrel{(\ref{TGS38})}{=}&-\frac{1}{2}\left( \alpha_{11}^2+2 \alpha_{12}^2+\alpha_{22}^2\right)=-\frac{11649}{190},\\
\label{IH7b}
E^{(2)}&\stackrel{(\ref{TGSSS2})}{=}& B_{sat}\,M^{(2)}=-\frac{34947}{190}
\;,
\end{eqnarray}
and, finally,
\begin{equation}\label{IH8}
 \chi_0\stackrel{(\ref{TGSSS4d})}{=}\frac{N}{B_{sat}+\frac{2 N^2}{\alpha_{11}^2+2 \alpha_{12}^2+\alpha_{22}^2}}=\frac{7766}{4643}
 \approx 1.67263
 \;.
\end{equation}
To check the latter result we have numerically calculated $20$ three-dimensional ground states for $B=0,\ldots,3$ and fitted
the corresponding function $M(B)$ by a polynomial. See Figure \ref{FIGGS4} for an example of the ground state where the
uniform coupling $\gamma$ is slightly above the critical value $\gamma_0$.
The slope of $M(B)$ at the saturation point fits very well
to the analytically determined saturation susceptibility $\chi_0$ according to Eq.~(\ref{IH8}), see
Figure \ref{FIGEX4}.

%%%%%%%%%%%%%%%%%%%%%%%%%%%%%%%%%%%%%%%%%%%%%%%%%%%%%%%%%%%%%%%%%%%%%%%%%%%%%%%%%%%%%%%%%%%%%%%%%%%%%%%%%%%%%%%%%%%%%%%%%%%%%%%%%%%%%%%%%%
\section{Summary and Outlook}\label{sec:SO}
%%%%%%%%%%%%%%%%%%%%%%%%%%%%%%%%%%%%%%%%%%%%%%%%%%%%%%%%%%%%%%%%%%%%%%%%%%%%%%%%%%%%%%%%%%%%%%%%%%%%%%%%%%%%%%%%%%%%%%%%%%%%%%%%%%%%%%%%%

We have applied the theory of ground states published three years ago to the problem of analytically describing the
behaviour of a spin system close to the saturation field where numerical calculations are difficult.
In particular, we have characterized the form of the ``spin umbrella"
in lowest order of its spread in terms of certain eigenvectors of the dressed $J$-matrix at the saturation point. Moreover,
we have derived simple expressions for the saturation susceptibility. This analysis has been performed for (locally) parabolic
systems and for non-parabolic systems with two- or three-dimensional ground states close to the saturation field and confirmed by
means of four examples using computer-algebraic software. We used the method of perturbation series up to fourth order for non-parabolic systems;
for the next interesting physical quantity, the slope of the saturation susceptibility, we would have to extend the perturbation series
up to the sixth order, which is possible in principle, but considerably more difficult. In view of the examples considered in this paper we
conjecture that the slope  of the saturation susceptibility will be negative for non-parabolic systems, as it is schematically indicated in Figure
\ref{FIGBM1}.

Although we think that our case distinction is complete for ``standard systems" it remains an open problem to extend the present theory
to those systems where the dimension $m$ of the ground states is larger than three and hence exceeds the domain of physically possible states.

%%%%%%%%%%%%%%%%%%%%%%%%%%%%%%%%%%%%%%%%%%%%%%%%%%%%%%%%%%%%%%%%%%%%%%%%%%%%%%%%%%%%%%%%%%%%%%%%%%%%%%%%%%%%%%%%%%%%%%%%%%%%%%%%%%%%%%%%%%
%\section*{Acknowledgment}
%%%%%%%%%%%%%%%%%%%%%%%%%%%%%%%%%%%%%%%%%%%%%%%%%%%%%%%%%%%%%%%%%%%%%%%%%%%%%%%%%%%%%%%%%%%%%%%%%%%%%%%%%%%%%%%%%%%%%%%%%%%%%%%%%%%%%%%%%

%\section*{References}

\end{document}